# Signatures of Reconnection and a Split Heliospheric Tail in High-Energy Energetic Neutral Atoms

M. Kornbleuth,[1*] M. Opher,[1] J. F. Drake,[2,3] M. Swisdak,[3] Zhiyu Yin,[2,3] K. Dialynas,[4] Y. Chen,[5] J. Giacalone,[6] J. M. Sokół,[7] M. Gkioulidou,[8] I. Baliukin,[9] V. Izmodenov,[9,10] G. P. Zank[11,12]

[1]Astronomy Department, Boston University, Boston, MA 02215, USA
[2] Department of Physics and Institute for Physical Science and Technology, University of Maryland, College Park, MD 20742, USA
[3] IREAP, University of Maryland, College Park, MD 20742, USA
[4] Center for Space Research and Technology, Academy of Athens, Athens 10679, Greece
[5] University of Michigan, Ann Arbor, MI 48109, USA
[6] Lunar & Planetary Laboratory, University of Arizona, Tucson, AZ 85721, USA
[7] Southwest Research Institute, P.O Drawer 28510, San Antonio, TX 78228, USA
[8] Applied Physics Laboratory, Johns Hopkins University, Laurel, MD 20723, USA
[9] HSE University, Moscow, 101000, Russia
[10] Lomonosov Moscow State University, Moscow, 119991, Russia
[11] Center for Space Plasma and Aeronomic Research (CSPAR), University of Alabama in Huntsville, Huntsville, AL 35899, USA
[12] Department of Space Science, University of Alabama in Huntsville, Huntsville, AL 35899, USA

*Correspondence to: Marc Kornbleuth, kmarc@bu.edu

## ABSTRACT

The shape of the heliosphere, regarded as comet-like since the 1960s, has recently been the subject of intense debate in the last decade. There is disagreement whether the heliospheric tail extends to ~10,000 au in a comet-like shape or if it is short (~ 400 au) with a split. Energetic neutral atom (ENA) maps from Cassini/INCA at energies from 5.2 to 13.5 keV revealed a global structure extending from the nose to the heliospheric tail known as the Belt whose origin has remained largely unexplored. Here, we use a state-of-the-art multi-ion magnetohydrodynamic (MHD) model and a novel reconnection simulation to establish that the Belt structure is consistent with a split tail heliosphere but not with a comet-like heliosphere. In a split-tail heliosphere there is a region of low- $\beta$ (ratio of thermal to magnetic pressure) in the downwind direction close to the heliopause. Direct simulations of this region reveal that magnetic reconnection is strong and drives the energetic particles that produce the >5.2keV ENAs measured by INCA in the low latitude portion of the Belt. Since the comet-like heliosphere does not produce this low- $\beta$ region and the resultant reconnection-drive mechanism for the >5.2keV ENAs, our results indicate that the INCA observations are inconsistent with a comet-like heliosphere. Further, these simulations and analysis establish for the first time that magnetic reconnection in the complex magnetic fields, expected in astrospheres across the universe, are likely to be a source of energetic particles and radiation.

## 1. INTRODUCTION

For almost seven decades (L. Davis 1955; E. N. Parker 1961; W. I. Axford 1972), the shape of the heliosphere has been viewed as a comet-like tail extending to ~10,000 au. However, recently there has been a revisiting of this paradigm and an active debate on the shape of the heliosphere (S. M. Krimigis et al. 2009; J. Kleimann et al. 2022; M. Opher et al. 2023). Numerical modeling studies have suggested that the shape may have a long, comet-like tail (V. B. Baranov & Y. G. Malama 1993; N. V. Pogorelov et al. 2015; V. V. Izmodenov & D. B. Alexashov 2015) or a short, split-tail resembling a croissant (Fig. 1) (M. Opher et al. 2015; M. Opher et al. 2020). There is consensus that the solar magnetic field plays a critical role in shaping the heliosphere because it confines the plasma in the heliosheath (the region of shocked solar wind between the termination shock and the heliopause) in two heliospheric jets (M. Opher et al. 2015; J. F. Drake et al. 2015; G. Yu 1974) in both a comet-like and split-tail heliosphere (V. V. Izmodenov & D. B. Alexashov 2015; M. Kornbleuth et al. 2021a).

The difference between the comet-like and split-tail heliosphere models is the strong diversion of plasma flow from low to high latitudes in the split-tail heliosphere (M. Kornbleuth et al. 2021a) as the radial flows approach the tailward heliopause approximately 350 au from the Sun. Within the models, these flows down-tail are only minimally present in the comet-like heliosphere. As we show below, these flows in the heliotail form a region where the magnetic pressure ($P_B$) exceeds the thermal pressure ($P_T$), resulting in a low-β plasma ($\beta=P_T/P_B$) not present in the comet-like heliosphere model. In the split-tail heliosphere model, magnetic reconnection likely occurs in this region.

Energetic Neutral Atoms (ENA) provide an indirect method for observing the heliosphere. They are products of charge exchange between solar wind ions and the interstellar neutral hydrogen gas. The IBEX-Hi instrument (H. O. Funsten et al. 2009) from the Interstellar Boundary Explorer (IBEX) mission (D. J. McComas et al. 2009) and the Ion and Neutral Camera (INCA) from the Cassini mission (S. M. Krimigis et al. 2009, K. Dialynas et al. 2013a) have been critical in producing global ENA images. The solar wind plasma at large distances downstream of the termination shock consists of a cold (~eV) and a hot suprathermal component (~keV). A significant constraint of ENA images is that the keV ions that produce them are quickly depleted as they charge exchange, limiting how far one can "see" down the heliospheric tail.

There have been attempts to deduce the structure of the heliosphere with ENA observations. These attempts were based on the assumption that the hot (~keV) ions in the heliosheath are only heated and accelerated at the termination shock (where the solar wind becomes subsonic). Under this assumption, using ENA observations of the globally distributed flux believed to originate from within the heliosheath, it had been concluded that observations with IBEX-Hi from 0.5-6 keV can only probe heliotail distances out to 300-400 au (D. B. Reisenfeld et al. 2021). ENAs at higher energies (5.2-55keV), as observed by INCA, are able to see farther down the tail due to their larger charge exchange mean free path (K. Dialynas et al. 2017). Empirical observations from INCA suggested that the heliosphere may have a bubble-like shape (S. M. Krimigis et al. 2009; K. Dialynas et al. 2017). High-energy ENA modeling (M. Kornbleuth et al. 2023a) -- up to 80 keV -- found that the highest INCA channels only extended the ENA distances to approximately 100 au more than the IBEX-Hi energies, which was insufficient for identifying the heliospheric shape considering different models of the heliosphere only have differences that are significant >400 au from the Sun (M. Kornbleuth et al. 2021a, 2021b). However, these conclusions were based on the assumption that ions are only accelerated at the termination shock and not elsewhere in the heliosheath. In-situ (Voyager) and remote sensing observations (IBEX and INCA) (e.g. K. Dialynas et al. 2020) seem to indicate that further acceleration inside the HS is in fact required (M. Gkioulidou et al. 2022). To "see" further down the tail and capture the full structure of the heliosphere (M. Kornbleuth et al. 2023a) requires either ENA observations at higher energies, such as those by the recently launched IMAP mission that will reach up to ~300 keV (D. J. McComas et al. 2025), or the existence of acceleration processes deep in the heliosheath. Here, we show that cold (~eV) ions of interstellar origin that are embedded in the solar wind plasma can be accelerated to keV energies in the heliotail via magnetic reconnection in the sector region as it piles up ahead of the heliopause. The ENAs from these energized ions reveal information on the shape of the heliosphere using currently available observations.

One prominent feature of ENA observations from Cassini/INCA is the Belt (Fig. 1) (S. M. Krimigis et al. 2009). The Belt is a feature of enhanced ENA flux that spans both the upwind and downwind hemispheres of the heliosphere. It covers a full-width half-maximum ranging from 60° to 100° in longitude across the sky (K. Dialynas et al. 2013a) and is produced by charge-exchange interactions between ions and neutrals inside the heliosheath (K. Dialynas et al. 2017). As seen in Fig. 2, the Belt passes directly through the low-latitude heliotail in the 5.2-13.5 keV energy band (central energy of 8.38 keV). High-latitude enhancements of ENA flux at high energies (>2 keV) were predicted due to the presence of fast, hot wind at high latitudes (D. J. McComas et al. 2013; E. J. Zirnstein et al. 2016a) and collimation of the plasma by the solar magnetic field (M. Kornbleuth et al. 2020). The source of the low-latitude enhancement in the heliotail has not been established.

The Belt is an energy-dependent structure in ENA maps (K. Dialynas et al. 2013a, 2017; J. M. Sokół et al. 2025). As seen in Fig. 2, the low-latitude heliotail enhancement of the Belt, herein referred to as the Buckle (spanning approximately from -30° to 30° in ecliptic latitude, and from 60° to 110° in ecliptic longitude), is not seen in IBEX-Hi observations in the 3.13-6 keV energy band (central energy of 4.29 keV). Likewise, the Buckle gradually fades in INCA observations in the 13.5-24 keV energy band (central energy of 18.00 keV), disappearing at higher energies. The narrow energy range of the Buckle may be evidence of an energization mechanism in the down-tail heliosheath. ENA models using only ion acceleration at the termination shock (M. Kornbleuth et al. 2023a) and adiabatic energy change in the heliosheath (I. Baliukin et al. 2023) have been unable to replicate the emergence and disappearance of the Buckle. Models indicate that the Belt must originate from a process beyond the termination shock. Thus far, no ENA model of the heliosphere has been able to replicate the full spatial structure and energy dependence of the Belt.

The solar wind at large distances from the Sun is composed of a cold thermal solar wind (eV) and a hot component called pick-up ions (PUIs). These PUIs are suprathermal (keV) ions that originate from the streaming neutral H from the interstellar medium which is ionized near the Sun and "picked up" by the solar magnetic field. In the heliosheath, models show that the PUIs dominate the thermal pressure. However, they are depleted by charge exchange as they penetrate further into the heliosheath. Previous multi-ion magnetohydrodynamic (MHD) modeling (M. Opher et al. 2020) considered the cold thermal solar wind ions and the PUIs as separate fluids. As PUIs charge exchange and leave the system, the heliosheath is cooled and the heliosphere deflates (M. Opher et al. 2020). As the PUIs are depleted in the heliotail, new ions are created in the heliosheath via charge exchange. This new population of heliosheath-created ions has an energy of ~0.1 keV. We show in our model that the depletion of the hot PUIs and the emergence of heliosheath-created ions, coupled with the compression of the solar magnetic field in the heliosheath, leads to the formation of a low-β region in the tail (Fig. 3).

The new population of heliosheath-created ions is modeled using a newly developed state-of-the-art global multi-ion MHD model. Inclusion of this new population is critical, because it allows us to model the creation of colder ions (eV) present in the heliosheath following the charge exchange depletion of the suprathermal ions accelerated at the termination shock. We describe in Materials and Methods why such an extension is critical to capturing the full extent of the low-β region in the tail. We argue that near the heliopause in the tail direction the high Alfven speed and low-β environment is likely to lead to magnetic reconnection and particle energization. To describe reconnection and associated particle acceleration, we use the *kglobal* model (J. F. Drake et al. 2019; H. Arnold et al. 2019, 2021) which is a fluid/particle hybrid model that has been recently upgraded to include particle ions (Zhiyu Yin et al. 2024a,b).

The *kglobal* model is the first model that is capable of calculating the extreme energization of ions produced during reconnection (M. Desai et al. 2025).

In Section 2, we provide an overview of our model and methods, with future details presented in the Appendix. In Section 3, we present our results and demonstrate that the 5.2-13.5 keV ENAs that constitute the Buckle, as measured by Cassini/INCA, are an observational signature of a split-tail heliosphere (Fig. 1). This is the first work that reproduces both the IBEX-Hi and INCA ENA fluxes and morphology within the same model.

## 2. METHODS

### 2.1 MHD Model

Our MHD model has three proton components (thermal solar wind, PUIs created in the supersonic solar wind, and heliosheath-created ions) that interact through charge exchange with neutral hydrogen atoms. The inner boundary of our domain is a sphere at 30 au and the outer boundary is at x = ±1500 au, y = ± 2000 au, z = ±2000 au. Within the supersonic solar wind in the upwind hemisphere, we have a grid resolution of 1 au, and in the heliosheath out to 400 au from the Sun in the heliospheric tail we have a resolution of 2 au. Further down the heliosphere out to 600 au we have 4 au resolution, and beyond we have 8 au resolution.

For the solar wind plasma flow at the inner boundary we use 22-year averaged solar cycle conditions (1995-2017) (V. V. Izmodenov & D. B. Alexashov 2020). This profile takes into account heliolatitudinal variations of the solar wind speed and density, and the temperature is related to the speed via an assumed constant Mach number of 1 au for all latitudes. We assume longitudinal symmetry. The solar wind conditions are based on a combination of OMNI 2 data, interplanetary scintillation observations, and SOHO/SWAN full-sky maps of backscattered Lyα intensities. This profile was previously used in a comparison of the split and long-tail heliospheres (M. Kornbleuth et al. 2021a).

We partition the solar wind plasma at our inner boundary by assuming constant density and energy fractions for the thermal and PUI populations for all latitudes. These fractions are derived from inner boundary conditions used in previous multi-ion MHD modeling (M. Opher et al. 2020) which are based on New Horizons predictions. We assume that 90% of the solar wind plasma density is composed of thermal solar wind ions, while 10% is composed of PUIs at 30 au. For the energy fractions of the plasma, we assume that the thermal energy components of the thermal solar wind ions and PUIs are 2% and 98%, respectively. There is an ad-hoc heating term applied to the PUIs in the supersonic solar wind to bring their temperature to $10^7$ K upstream of the termination shock (M. Opher et al. 2020) as predicted by New Horizons (D. J. McComas et al. 2021). The PUIs at the inner boundary are chosen to have the same speed as the solar wind, but beyond the inner boundary the two species are not considered to be co-moving. For the heliosheath-created ions, we are required to include them at the inner boundary within our modeling, but we assign negligible density ($10^{-5}$ $n_{PUI}$) and pressure ($10^{-5}$ $P_{PUI}$) such that they do not affect the solution in the supersonic solar wind. This population is also considered to have the same speed as the thermal solar wind ions and the PUIs at the inner boundary and is also not considered to be co-moving beyond it.

The magnetic field is given by the Parker spiral magnetic field with $B_{SW} = 7.17x10^{-3}$ nT at the solar equator. We use a unipolar configuration for the solar magnetic field to minimize artificial reconnection effects, such as in the heliospheric current sheet. This description captures the topology of the solar magnetic field lines, but there is only a single polarity of the magnetic field throughout the domain. In our simulation, we assume that the magnetic axis is aligned with the solar rotation axis.

As shown in Fig. 3, in the quasi-stagnation region within the heliospheric tail of the split-tail heliosphere $P_B$ increases sufficiently such that it exceeds $P_T$. This produces a region that is characterized by $\beta < 1$ in our MHD model. Models generally produce a high-β region in the nose direction (V. V. Izmodenov & D. B. Alexashov 2015), except in the region near the heliopause where models predict a strong pile-up of the solar magnetic field exceeding Voyager observations (A. T. Michael et al. 2018). Voyager observations (L. F. Burlaga & N. F. Ness 2012) show a roughly constant solar magnetic field intensity of around ~0.1 nT until the radial velocity of the solar wind plasma begins to decrease as Voyager 1 approaches the heliopause. In contrast, MHD models treating the solar magnetic field as either dipolar or unipolar see a steady increase of the solar magnetic field from the termination shock until the heliopause. Since it eliminates numerical dissipation of the magnetic field, the unipolar model overpredicts the solar magnetic field in the heliosheath compared to a dipolar magnetic field, but by no more than a factor of ~1.5 in the Voyager 1 direction (A. T. Michael et al. 2018). The MHD model included in this work uses a unipolar magnetic field, with the thermal pressure of the plasma as approximately $7 \times 10^{-15}$ dynes/cm$^2$ and a solar magnetic field strength of 0.25 nT in the low-β region. If we account for the difference in the solar magnetic field magnitude between the unipolar and dipolar cases, in the dipolar case the solar magnetic field intensity in the region of interest would be 0.17 nT. For the unipolar case, we obtain $\beta \sim 0.3$, whereas for the dipolar case we still obtain a low-β region, with $\beta \sim 0.65$.

For the interstellar plasma, we use the conditions from our previous multi-ion modeling (M. Opher et al. 2020) and assume: $v_{ISM}$ = 26.4 km/s, $n_{ISM}$ = 0.06 cm$^{-3}$, $T_{ISM}$= 6519 K. The number density of H atoms in the interstellar medium is $n_H = 0.18$ cm$^{-3}$, the velocity and temperature are the same as for the interstellar plasma. The coordinate system is such that the Z-axis is parallel to the solar rotation axis, the X-axis is 5° above the direction of interstellar flow with the Y-axis completing the right-handed coordinate system. The interstellar magnetic field ($B_{ISM}$) direction and strength are based on the IBEX-ribbon derived direction (E. J. Zirnstein et al. 2016b) and are in reasonable agreement with the Voyager 1 and 2 magnetic field observations outside of the heliopause (M. Opher et al. 2020), respectively. The strength of the interstellar magnetic field ($B_{ISM}$) in the model is 3.2μG and the orientation of $B_{ISM}$ is -34°.62 and 47°.3 in ecliptic latitude and longitude, respectively. We note that recent works (E. Powell et al. 2024; M. Kornbleuth et al. 2024) have suggested a less-inclined $B_{ISM}$ direction relative to the interstellar flow in the hydrogen deflection plane (R. Lallement et al. 2005, 2010). Considering the uncertainty of $B_{ISM}$ in the community, here we elected to use the best configuration from previous multi-ion modeling (M. Opher et al. 2020).

### 2.2 ENA Model

We update our previous ENA model (M. Kornbleuth et al. 2023b). We extend the model to consider a multi-ion MHD plasma and to include the effects of reconnection. Details of our ENA model are included in the Appendix.

## 3. RESULTS

We simulate the heliosphere with our newly extended MHD model. It describes the population of ions created in the heliosheath via charge exchange as a separate fluid (besides the thermal and suprathermal PUI fluids) (M. Opher et al. 2020) and treats the incoming neutrals atoms from the interstellar medium atoms kinetically (see Appendix).

### 3.1 A Low-β Region in the Heliotail

In the downwind portion of the split-tail heliosphere there is a quasi-stagnation point near the heliopause between the two lobes (Figs. 3A and 3B). As the plasma flows from the termination shock into the heliotail at low latitudes it encounters the heliopause and is deflected to higher latitudes (M. Opher et al. 2015, M. Kornbleuth et al. 2021a). This deflection leads to a quasi-stagnation region around 350 au from the Sun (Fig. 3) and a pile-up of the solar magnetic field (Figs. 3D and 3E). Based on Voyager observations, the heliosheath in the upwind direction is found to be a high-$\beta$ plasma since the PUIs (~keV) dominate the thermal pressure which is an order of magnitude larger than the magnetic pressure. This assumption breaks down in the tail at distances where the PUIs created in the supersonic solar wind are depleted due to charge exchange. As these PUIs are depleted, they are replaced with lower energy ions (eV) in the heliosheath, effectively cooling the solar wind plasma. A low-$\beta$ region is created from the pileup of the magnetic field and the low thermal pressure of the cooled plasma.

As shown in Fig. 3, this low-$\beta$ region is unique to the split-tail heliosphere (Figs. 3K and 3L). Within a comet-like model of the heliosphere, the tailward flows of plasma are uninterrupted for large distances from the Sun and there is only minimal deflection at low latitudes (Fig. 3C). To quantify this, we compare two single-ion MHD models with different heliospheric shapes. In the split-tail model the low-$\beta$ region begins approximately 475 au from the Sun and extends for 50 au until the heliopause (Figs. 3B and 3E). In contrast, in the long-tail model we find $\beta > 1$ as far from the Sun as 750 au (Figs. 3C and 3F). The single-ion MHD models in Fig. 3 use the same boundary conditions for the solar and interstellar winds, where the split-tail case is the BU model and the long-tail case is the Moscow model (M. Kornbleuth et al. 2021a).

### 3.2 Magnetic Reconnection in the Sector Region in the Low-$\beta$ Region

The heliospheric current sheet separates regions of opposite magnetic polarity. Due to the flapping of the sheet a latitudinal zone called the sector region is produced in which radial trajectories pass between magnetic polarities. The latitudinal extent of the sector region varies with solar cycle, but always persists at low enough latitudes. As the solar wind plasma is cooled and slowed ahead of the tail heliopause, the heliospheric current sheet embedded within the sector region is compressed.

The sector region can undergo magnetic reconnection, which may be a potential source of ion acceleration. J. F. Drake et al. (2010) and M. Opher et al. (2011) argued that reconnection should occur in the nose direction as the spacing between the sector regions drops to kinetic scales due to the slowing of the plasma flows. Global MHD models cannot capture the sector region due to numerical dissipation (M. Opher et al. 2011; N. V. Pogorelov et al. 2013). In the low-latitude heliospheric tail the plasma flows ahead of the heliopause in the split-tail heliosphere slow down, as at the nose. These flows are diverted to higher latitudes due to the collimation of the plasma by the solar magnetic field. At low latitudes, the sector is compressed within the quasi-stagnation region. Reconnection should be occurring in the compressed sector region as well. As we show in Fig. 3G-L, the magnetic field compression only occurs in the split-tail heliosphere. We chose to model the magnetic field as unipolar within our MHD model to avoid numerical dissipation (see Appendix). We determine the region where the sector compression occurs by identifying where the build-up in the magnetic field intensity in the heliosheath occurs. This region directly correlates with the low-β region within our MHD model.

To evaluate the amount of magnetic energy available for accelerating ions within this low-β region (Fig. 4), we consider the quantity $mV_A^2$, where m is the proton mass and $V_A$ is the Alfvén speed. This quantity measures the magnetic energy per particle available during reconnection (T. D. Phan et al. 2013, 2014). It should be compared to the local thermal energy of the heliosheath-created ions, which dominates the plasma by density in the low-β region, to evaluate when reconnection can produce significant energization. The maximum value of $mV_A^2$ in the split-tail model is more than a factor of two higher than in the comet-like model of the heliosphere (Fig. 3). This maximum coincides with the low-β region of the split-tail model and, moreover, the volume occupied by the region of high $mV_A^2$ in the split tail model greatly exceeds that of the comet-like model. This is a result of the build-up of the solar magnetic field ahead of the heliopause.

The plasma in the low-β region is largely composed of ions created in the heliosheath via charge exchange. As shown in Fig. 4, these heliosheath-created ions initially have temperatures around 0.1 keV near the termination shock. The temperatures decrease to approximately 0.01-0.02 keV in the low-β region. By comparison, $mV_A^2$ in this region is approximately 0.25 keV. The significantly larger $mV_A^2$ indicates a large amount of available magnetic energy that is capable of accelerating eV ions to keV energies in the presence of magnetic reconnection. Recent observations of reconnection (M. Desai et al. 2025) by the Parker Solar Probe (PSP) in the near-sun heliospheric current sheet (HCS) have revealed that protons can reach energies up to $10^3$ times $mV_A^2$. Direct simulations of the PSP event with the *kglobal* model reproduced this extreme energization (M. Desai et al. 2025).

Based on the recent observations at the HCS, we might expect that protons in the heliotail can be accelerated up to energies approaching 250 keV ($10^3$ times $mV_A^2$). On the other hand, such extreme energization is not consistent with the INCA observations, which revealed ENAs at 8.38 keV from the low latitude heliotail but not ENAs at 18.0 keV. Reconnection modeling has identified the ambient guide field $B_g$ (the magnetic field component perpendicular to the plane of reconnection) as a key control parameter (H. Arnold et al. 2021; Zhiyu Yin et al. 2024b; J. F. Drake et al. 2006). Thus, our strategy is to carry out a series of simulations with different guide fields $B_g$ and initial conditions based on the local parameters in the heliotail from our global MHD model (temperatures of 0.0156 keV and $mV_A^2$ = 0.25 keV). The simulations are carried out with the *kglobal* model, which combines MHD and particle dynamics to describe reconnection and particle energization in macro-systems (J. F. Drake et al. 2019; H. Arnold et al. 2019, 2021). Because the model includes the feedback of energetic particles on the MHD dynamics (while conserving the total energy of the system), the model is a major advance compared with test particle treatments in MHD models. The model originally only included particle electrons but has recently been upgraded to include particle protons (Zhiyu Yin et al. 2024a,b). The initial temperature of 0.0156 keV, on average, reflects the heliosheath-created ions that dominate the plasma by number density in the low-β region. In Fig. 4 we show the results of the proton spectra (averaged over the entire simulation domain) from reconnection simulations using *kglobal* with guide fields of $B_g/B_0$ = 0.25, 0.5, and 0.75, where $B_g$ is the magnitude of the guide field and $B_0$ is the magnitude of the reconnecting magnetic field ($B_0$ ~ 0.3 nT in the low-β region). We treat $B_g/B_0$ as a free parameter because we are unable to resolve the reconnection plane within our model. Further details on the setup of these simulations can be found in Zhiyu Yin et al. (2024b).

We then create ENA maps including the reconnection-derived proton spectrum in the region of the heliotail where the available magnetic energy exceeds the thermal energy ($E_{th}$ = $k_BT$) of the plasma (e.g., $mV_A^2/E_{th} > 1$). This specifies locations where reconnection is likely to produce particle energization in the heliotail. The accelerated ion spectrum in this region produces an energy-dependent enhancement of ENA flux. As expected, the hard spectra resulting from the weaker guide field simulations ($B_g/B_0 = 0.25$ and 0.5) produce a strong signal in the low latitude 18.0 keV ENAs that is inconsistent with the INCA observations (see Appendix). In contrast, the data from the $B_g/B_0 = 0.75$ simulation produces ENA enhancements in the 8.38 keV energy band at low latitudes, but no significant enhancement in the 18.0 keV channel (Fig. 2). These ENA profiles qualitatively agree with both the IBEX-Hi and INCA ENA observations of the heliotail. Our results highlight the importance of reconnection in the heliotail and may indicate observational evidence of reconnection occurring in the heliosheath. The results also offer support for the split tail structure of the heliosphere since models that produce the comet-like structure do not produce a high Alfven speed, low-$\beta$ region in the heliotail.

## 4. DISCUSSION

This is the first time that ENA modeling has been able to replicate the full morphology of the Belt in the heliotail at INCA energies. Additionally, this work is the first to replicate the ENA morphology of the transition from IBEX-Hi ENA observations (< 6 keV) to INCA ENA observations (> 5.2 keV). We show that the inclusion of reconnection in the sector region of the heliosheath ahead of the heliopause in the tail can replicate the Buckle. There is a large volume of the heliotail where $mV_A^2$ is large (~ 0.25-0.3keV) in the split-tail model. In contrast, $mV_A^2$ is much smaller in the comet-like model and is more spatially localized. As a consequence, ion energization and ENA production are expected to be much stronger in the split tail system compared with a comet-like system. There is a region beyond the heliopause in the split-tail heliosphere where there is mixing of interstellar and solar wind plasma along reconnected field lines that join the interstellar and solar magnetic fields. We do not consider reconnection in this mixing region because the piled-up sector region is absent there. In future observations by IMAP, which will be able to observe ENAs up to 300 keV, we will be able to further probe particle energization in this region.

The prospect of reconnection in the nose of the heliosphere due to the compression of the sector region has been explored by several authors (J. F. Drake et al. 2010; M. Opher et al. 2011). However, it is not clear from existing observations (J. D. Richardson et al. 2013, 2021) if reconnection is occurring in the nose. Considering the ENA spectra from the nose and tail directions in INCA observations are very similar (S. M. Krimigis et al. 2009; Dialynas et al. 2013, 2017), reconnection explaining ENA observations in the tail suggests possible applicability in the nose as well. A more detailed reconnection modeling effort on this topic can be carried out now that *kglobal* is fully operational. We do note in Fig. 2 that our model underpredicts the ENA flux at the nose, which could be potentially resolved by reconnection. Yet due to the uncertainty of the role of reconnection in the nose, we do not include reconnection driven energetic protons within this region in our ENA modeling for this work.

While the *kglobal* and MHD results account for most of the INCA observations, the ENA results here support the findings of a discrepancy between high energy ENA observations and ENA modeling (M. Gkioulidou et al. 2022; M. Kornbleuth et al. 2023b; J. Giacalone et al. 2025), specifically a noticeable gap between ENA models and data from 4-20 keV. The gap has its largest discrepancy in the 8.38 keV energy band when only ion acceleration processes at the termination shock are treated. K. Dialynas et al. (2019) also indicated that the >5.2 keV ENAs from the Belt serve as important indicators of the acceleration processes that the parent proton population undergoes inside the heliosheath. Analyses of ions and ENAs inside the heliosheath further highlighted the need for additional acceleration of ions beyond the termination shock (K. Scherer et al. 2022), suggesting that there is likely a global heliosheath-based acceleration of ions based on the omnidirectionality of the discrepancy. However, the discrepancy (M. Kornbleuth et al. 2023a,b) was shown to increase in the region of the heliotail where the Belt occurs. The data-to-model ratio at the 8.38 keV central energy for the Voyager 2 direction using ENA data averaged from 2009-2012 was 5.00. In the low-latitude tail where the belt occurs, the ratio was 18.39. This contrast indicated that an additional energy-dependent process in the heliotail is required to replicate ENA observations. We note that the averaged 2009-2012 data from INCA excludes 2012 data in the tail direction, which corresponds to a year with lower flux than the preceding year (K. Dialynas et al. 2017). Therefore, the ENA data from INCA averaged from 2009-2012 may be inflated by at least 15-20% (M. Kornbleuth et al. 2023b). Magnetic reconnection in the heliotail helps to resolve the directional discrepancy between the model and the data in the low-latitude tail, but not entirely. It also appears to be the missing link in obtaining the ENA morphology in the tail at both IBEX-Hi (~0.5-6 keV) and INCA (>5.2 keV) energies.

Two notable limitations regarding the modeling aspect of this work include a lack of evolving solar wind conditions and the persistence of an omnidirectional gap between ENA modeling and observations. While time-dependent solar wind conditions are important for both IBEX (J. M. Sokół et al. 2021; D. J. McComas et al. 2024) and the INCA ENAs (K. Dialynas et al. 2017) with the Belt reducing in both intensity and width through the declining phase of SC23, it does not vanish. Though the width and behavior of the sector region will vary with solar cycle, regardless of the phase of the solar cycle the sector region will always be present at low latitudes where the Buckle is formed. Therefore, while solar wind effects may affect the amount of flux observed in the Buckle, our modeling investigation does not require a time-dependent model to prove the role of reconnection on the formation of the enhanced ENA emission in the Buckle.

For future investigation, we will explore the effect of changing solar cycle on the reconnection mechanism to probe whether the Buckle remains throughout the cycle or undergoes periods where it disappears due to unfavorable conditions for magnetic reconnection. Additionally, while a gap persists between ENA modeling and observations at high energies (M. Kornbleuth et al. 2023b), that is not the focus of this study since the Buckle is confined to the heliotail. Therefore, its source process differs from the source process contributing to the omnidirectional gap.

A potential limitation regarding the INCA observations that could have implications on our conclusions is regarding the point spread function (PSF) of the observations. The 8.38 keV energy band of INCA, which is limited on the low end by the minimum penetration energy of hydrogen atoms through the ~10μg/cm2 start foil in the entrance slit. Scattering in the start foil for those H atoms that are transmitted is significant, resulting in a broad angular point spread function PSF. K. Dialynas et al. (2013b) showed that the full width half maximum (FWHM) at 36 keV was 11° in elevation angle and 9° in azimuthal angle. Considering the FWHM should be wider over the ~5.2 – 13.5 keV bandwidth, on the order of 30° or a bit more based on Meyer thin foil scattering theory (L. Meyer 1971), it does lead to an uncertainty in the measurements that can affect our conclusions. The gap in emission at the equator is of the order of 20 to 30 degrees in latitudinal extent as measured at 4.29 keV by IBEX and at 18 keV by INCA. This could lead to some difficulty in discerning a feature of a seemingly similar angular scale given that the FWHM at 8.38 keV should be wider than that of ~36 keV. However, here we should point out that, even though there is a gap in emission at the equator at the 18 keV channel, as shown in the paper, the morphology of the tail at the 35-55 keV INCA channel is very similar to the 5.2-13.5 keV channel (e.g. Fig 1. in K. Dialynas et al. 2013a), i.e., there is no discernible gap. The IMAP observations, which will provide enhanced statistics and significantly higher angular resolution, are necessary to address this potential limitation.

Our results support the split-tail shape of the heliosphere (M. Opher et al. 2020) and all models and observations (S. M. Krimigis et al. 2009; K. Dialynas et al. 2013a, 2017) that argue for a short tail on the scale of hundreds of au, instead of the traditional comet-like heliosphere. Determining the shape of the heliosphere is not only important for determining the physics within the heliosphere, but also may have important implications for our local radiation environment, such as galactic cosmic ray (GCR) transport into the heliosphere. Understanding how GCRs are filtered through a split, turbulent heliotail is therefore critical to properly constrain our understanding of GCRs in the heliosphere and for other astrospheres.

## Appendix

### A1. MHD MODEL: DESCRIPTION OF THE GOVERNING EQUATIONS

We treat the neutral H atoms in two separate ways: as fluids and kinetically. We first run our MHD model to steady state using a multi-fluid approach for the neutrals. Once a steady state MHD solution is obtained with a multi-fluid neutral treatment, we couple the MHD model to a kinetic neutral code, FLEKS (Y. Chen et al. 2023) (similar to the process of A. T. Michael et al. 2022). The neutral H atoms are described in a kinetic treatment with adaptive mesh refinement and particle splitting to improve the simulation efficiency. The charge exchange source terms are calculated in FLEKS based on the ion parameters for a given time and location and are passed back to the multi-ion MHD model to update the parameters of each ion species. This iterative process is conducted in a time-dependent mode such that the source terms are passed back to the MHD model every 0.05 years. Further details about the FLEKS kinetic neutral model and its MHD coupling can be found in Y. Chen et al. 2024.

The model assumes the "cold electron" approximation, i.e., that there are no suprathermal electrons. This is in agreement with the observations (J. D. Richardson et al. 2008). The number density of the thermal solar wind protons, PUIs created in the supersonic solar wind, and heliosheath-created ions are represented as $n_{SW}$, $n_{PUI}$, and $n_{HSI}$, respectively. From charge neutrality we have,

$$n_e = n_{SW} + n_{PUI} + n_{HSI} \tag{2}$$

We model the electrons as a separate fluid in our MHD model. Considering charge neutrality and their negligible momentum relative to the ion species, we only consider a separate pressure equation for the electrons. Details of the inclusion of electrons as a separate species can be found in E. S. Bair et al. (2025). The thermal pressure for each ion species is given by $p_i = n_i T_i k_B$, where the subscript *i* reflects each ion species. The separate treatment of the electrons is critical for fully resolving the low-β region of the heliotail compared to single-ion MHD models. Single-ion models by necessity assign a thermal temperature for the electrons equal to the total plasma temperature. This leads to an overprediction of the electron temperature that inflates the thermal pressure within the heliotail. This yields a significantly smaller low-β region in the heliotail as can be seen in Fig. 3 when comparing the multi-ion model (panel A) and single-ion model (panel B) for the split-tail heliosphere.

We expand the multi-fluid set of equations (M. Opher et al. 2020; A. Glocer et al. 2009; G. Toth et al. 2012) for the ions to include source terms due to charge exchange,

$$\frac{\partial \rho_{SW}}{\partial t} + \vec{\nabla} \cdot (\rho_{SW} \vec{u}_{SW}) = S_{\rho_{SW}} \tag{3}$$

$$\frac{\partial \rho_{PUI}}{\partial t} + \vec{\nabla} \cdot (\rho_{PUI} \vec{u}_{PUI}) = S_{\rho_{PUI}} \tag{4}$$

$$\frac{\partial \rho_{HSI}}{\partial t} + \vec{\nabla} \cdot (\rho_{HSI} \vec{u}_{HSI}) = S_{\rho_{HSI}} \tag{5}$$

$$\frac{\partial(\rho_{SW}\vec{u}_{SW})}{\partial t} + \vec{\nabla}\cdot\left(\rho_{SW}\vec{u}_{SW}\vec{u}_{SW} + p_{SW}\overleftrightarrow{I}\right) - \frac{\rho_{SW}}{m_p}(\vec{u}_{SW} - \vec{u}_{+})\times\vec{B} - \frac{\rho_{SW}}{n_e e}\vec{J}\times\vec{B} = S_{M_{SW}} \tag{6}$$

$$\frac{\partial(\rho_{PUI}\vec{u}_{PUI})}{\partial t} + \vec{\nabla}\cdot\left(\rho_{PUI}\vec{u}_{PUI}\vec{u}_{PUI} + p_{PUI}\overleftrightarrow{I}\right) - \frac{\rho_{PUI}}{m_p}(\vec{u}_{PUI} - \vec{u}_{+})\times\vec{B} - \frac{\rho_{PUI}}{n_e e}\vec{J}\times\vec{B} = S_{M_{PUI}} \tag{7}$$

$$\frac{\partial(\rho_{HSI}\vec{u}_{HSI})}{\partial t} + \vec{\nabla}\cdot\left(\rho_{HSI}\vec{u}_{HSI}\vec{u}_{HSI} + p_{HSI}\overleftrightarrow{I}\right) - \frac{\rho_{HSI}}{m_p}(\vec{u}_{HSI} - \vec{u}_{+})\times\vec{B} - \frac{\rho_{HSI}}{n_e e}\vec{J}\times\vec{B} = S_{M_{HSI}} \tag{8}$$

$$\frac{\partial\mathcal{E}_{SW}}{\partial t} + \vec{\nabla}\cdot[(\mathcal{E}_{SW} + p_{SW})\vec{u}_{SW})] - \frac{\rho_{SW}}{m_p}\vec{u}_{SW}\cdot(\vec{u}_{SW} - \vec{u}_{+})\times\vec{B} - \frac{\rho_{SW}}{n_e e}\vec{u}_{SW}\cdot\vec{J}\times\vec{B} = S_{\mathcal{E}_{SW}} \tag{9}$$

$$\frac{\partial\mathcal{E}_{PUI}}{\partial t} + \vec{\nabla}\cdot[(\mathcal{E}_{PUI} + p_{PUI})\vec{u}_{PUI})] - \frac{\rho_{PUI}}{m_p}\vec{u}_{PUI}\cdot(\vec{u}_{PUI} - \vec{u}_{+})\times\vec{B} - \frac{\rho_{PUI}}{n_e e}\vec{u}_{PUI}\cdot\vec{J}\times\vec{B} = S_{\mathcal{E}_{PUI}} + H, \tag{10}$$

$$\frac{\partial\mathcal{E}_{HSI}}{\partial t} + \vec{\nabla}\cdot[(\mathcal{E}_{HSI} + p_{HSI})\vec{u}_{HSI})] - \frac{\rho_{HSI}}{m_p}\vec{u}_{HSI}\cdot(\vec{u}_{HSI} - \vec{u}_{+})\times\vec{B} - \frac{\rho_{HSI}}{n_e e}\vec{u}_{HSI}\cdot\vec{J}\times\vec{B} = S_{\mathcal{E}_{HSI}}, \tag{11}$$

where $\vec{u}_{+} = \frac{\rho_{SW}\vec{u}_{SW} + \rho_{PUI}\vec{u}_{PUI} + \rho_{HSI}\vec{u}_{HSI}}{\rho_{SW} + \rho_{PUI} + \rho_{HSI}}$ is the charge-averaged ion velocity and the source terms, S represent the mass, momentum, and energy sources respectively due to charge exchange (R. L. McNutt et al. 1988). In equation 10, “H” reflects a PUI heating term to replicate New Horizons observations showing a heating of PUIs as a function of distance (M. Opher et al. 2020). Radiation pressure and gravity are assumed to perfectly cancel each other out. Ionization processes such as photoionization and electron-impact ionization are also neglected. These processes play a much lesser role than charge exchange at larger radii (R > 30 AU), especially in the upwind direction (J. M. Sokół 2019a,b; I. Kowalska-Leszczynska et al. 2022).

For the multi-fluid neutral treatment, the neutral H atoms are described as 4 different populations having the characteristics of different regions of the heliosphere (M. Opher et al. 2009). The four populations of neutral hydrogen atoms have different origins: atoms of interstellar origin represent population 4; population 1 is created by charge exchange in the region of interstellar space disturbed by the presence of the heliosphere, also known as the outer heliosheath, and populations 3 and 2 originate from the supersonic solar wind and the heliosheath, respectively. All four neutral fluid populations, distinguished by index "i", are described by separate systems of Euler equations with the corresponding source terms describing the ion-neutral charge exchange process,

$$\frac{\partial \rho_H(i)}{\partial t} + \vec{\nabla} \cdot (\rho_H \vec{u}_H) = S_{\rho_H}(i) \tag{12}$$

$$\frac{\partial \rho_H \vec{u}_H}{\partial t} + \vec{\nabla} \cdot \left(\rho_H \vec{u}_H \vec{u}_H + p_H \overleftrightarrow{I}\right) = S_{M_H}(i) \tag{13}$$

$$\frac{\partial \mathcal{E}_H}{\partial t} + \vec{\nabla} \cdot [(\mathcal{E}_H + p_H)\vec{u}_H)] = S_{\mathcal{E}_H}(i) \tag{14}$$

While the neutrals are treated kinetically, within FLEKS we also tag individual neutral particles with a marker depending on where they are created to track the evolution of different species.

**A1.1 MHD Model: Overview of Source Terms**

Here we describe our multi-fluid description of the neutrals, which allows for the interaction of ions and neutrals via charge exchange. In the supersonic solar wind (Region 3), the following charge-exchange processes occur,

$$p_0 + H_1 \rightarrow p_1 + H_3$$

$$p_0 + H_2 \rightarrow p_1 + H_3$$

$$p_0 + H_3 \rightarrow p_0 + H_3$$

$$p_0 + H_4 \rightarrow p_1 + H_3$$

and

$$p_1 + H_1 \rightarrow p_1 + H_3$$

$$p_1 + H_2 \rightarrow p_1 + H_3$$

$$p_1 + H_3 \rightarrow p_0 + H_3$$

$$p_1 + H_4 \rightarrow p_1 + H_3$$

and

$$p_2 + H_1 \rightarrow p_1 + H_3$$

$$p_2 + H_2 \rightarrow p_1 + H_3$$

$$p_2 + H_3 \rightarrow p_0 + H_3$$

$$p_2 + H_4 \rightarrow p_1 + H_3$$

where $p_0$ is the solar wind proton, $p_1$ the PUI created in the supersonic solar wind, $p_2$ is the heliosheath-created ion, and $H_1$, $H_2$, $H_3$, $H_4$ are, respectively, neutrals from populations 1, 2, 3 and 4. While $p_2$ is only created in the heliosheath and therefore should not be present in Region 3, for the purposes of our fluid description we are required to consider its presence. Therefore, we assign a negligible density and pressure to heliosheath-created ions in this region so that it does not affect our solution here.

Outside of Region 3, the following charge exchange processes occur in the heliosheath (Region 2),

$$p_0 + H_1 \rightarrow p_2 + H_2$$

$$p_0 + H_2 \rightarrow p_2 + H_2$$

$$p_0 + H_3 \rightarrow p_2 + H_2$$

$$p_0 + H_4 \rightarrow p_2 + H_2$$

and

$$p_1 + H_1 \rightarrow p_2 + H_2$$

$$p_1 + H_2 \rightarrow p_2 + H_2$$

$$p_1 + H_3 \rightarrow p_2 + H_2$$

$$p_1 + H_4 \rightarrow p_2 + H_2$$

and

$$p_2 + H_1 \rightarrow p_2 + H_2$$

$$p_2 + H_2 \rightarrow p_2 + H_2$$

$$p_2 + H_3 \rightarrow p_2 + H_2$$

$$p_2 + H_4 \rightarrow p_2 + H_2$$

Therefore, in Region 2 all charge exchange events result in the creation of the heliosheath-created ion species. Likewise, we also replicate this same process in the outer heliosheath (Region 1) to replicate newly created ions within this region,

$$p_0 + H_1 \rightarrow p_2 + H_1$$

$$p_0 + H_2 \rightarrow p_2 + H_1$$

$$p_0 + H_3 \rightarrow p_2 + H_1$$

$$p_0 + H_4 \rightarrow p_2 + H_1$$

and

$$p_1 + H_1 \rightarrow p_2 + H_1$$

$$p_1 + H_2 \rightarrow p_2 + H_1$$

$$p_1 + H_3 \rightarrow p_2 + H_1$$

$$p_1 + H_4 \rightarrow p_2 + H_1$$

and

$$p_2 + H_1 \rightarrow p_2 + H_1$$

$$p_2 + H_2 \rightarrow p_2 + H_1$$

$$p_2 + H_3 \rightarrow p_2 + H_1$$

$$p_2 + H_4 \rightarrow p_2 + H_1$$

Like in Region 3, we note that $p_1$ should not exist in Region 1 since these ions are restricted to the heliosphere. However, for the purposes of the fluid treatment of the ions we assign negligible density and pressure to $p_1$.

In the source terms the following terms appear, where the index "*i*" refers to each population of neutrals: 1, 2, 3, or 4. $U_{thSW}^2$ is the thermal speed of the solar wind, $U_{thPUI}^2$ the thermal speed of PUIs, and $U_{thHSI}^2$ the thermal speed of heliosheath-created ions:

$$U^*{}_{SW}(i) = \sqrt{\frac{4}{\pi}(w_{SW}^2 + w_H^2(i)) + (\Delta U_{SW-H}(i))^2}, U_{thSW}^2 = \frac{2k_B T_{SW}}{m_p}, U_{th}^2(i) = \frac{2k_B T_H(i)}{m_p},$$

$$U_{thPUI}^2 = \frac{2k_B T_{PUI}}{m_p}, U^*{}_{PUI}(i) = \sqrt{\frac{4}{\pi}(w_{PUI}^2 + w_H^2(i)) + (\Delta U_{PUI-H}(i))^2},$$

$$U_{thHSI}^2 = \frac{2k_B T_{HSI}}{m_p}, U^*{}_{HSI}(i) = \sqrt{\frac{4}{\pi}(w_{HSI}^2 + w_H^2(i)) + (\Delta U_{HSI-H}(i))^2},$$

$$\Delta U_{SW-H}(i) = |\vec{u}_H(i) - \vec{u}_{SW}|; \; \Delta U_{PUI-H}(i) = |\vec{u}_H(i) - \vec{u}_{PUI}|; \Delta U_{HSI-H}(i) = |\vec{u}_H(i) - \vec{u}_{HSI}|$$

$$U^*{}_{M-PUI}(i) = \sqrt{\frac{64}{9\pi}(w_{PUI}^2 + w_H^2(i)) + (\Delta U_{PUI-H}(i))^2},$$

$$U^*{}_{M-HSI}(i) = \sqrt{\frac{64}{9\pi}(w_{HSI}^2 + w_H^2(i)) + (\Delta U_{HSI-H}(i))^2},$$

$$U^*{}_{M-SW}(i) = \sqrt{\frac{64}{9\pi}(w_{SW}^2 + w_H^2(i)) + (\Delta U_{SW-H}(i))^2}.$$

The cross sections are from B. G. Lindsay & R. F. Stebbings (2005),

$$\sigma_{SW}(i) = (2.2835 \times 10^{-7} - 1.062 \times 10^{-8} \ln(U^*{}_{M-SW}(i) * 100))^2 \times 10^{-4}\, cm^2$$

$$\sigma_{NSW}(i) = (2.2835 \times 10^{-7} - 1.062 \times 10^{-8} \ln(U^*{}_{SW}(i) * 100))^2 \times 10^{-4}\, cm^2$$

$$\sigma_{PUI}(i) = (2.2835 \times 10^{-7} - 1.062 \times 10^{-8} \ln(U^*{}_{M-PUI}(i) * 100))^2 \times 10^{-4}\, cm^2$$

$$\sigma_{NPUI}(i) = (2.2835 \times 10^{-7} - 1.062 \times 10^{-8} \ln(U^*{}_{PUI}(i) * 100))^2 \times 10^{-4}\, cm^2$$

$$\sigma_{HSI}(i) = (2.2835 \times 10^{-7} - 1.062 \times 10^{-8} \ln(U^*{}_{M-HSI}(i) * 100))^2 \times 10^{-4}\, cm^2$$

$$\sigma_{NHSI}(i) = (2.2835 \times 10^{-7} - 1.062 \times 10^{-8} \ln(U^*{}_{HSI}(i) * 100))^2 \times 10^{-4}\, cm^2$$

**A1.2 MHD Model: Region 3 (Supersonic Solar Wind) Source Terms**

In Region 3, in the supersonic solar wind, the density source term for the solar wind protons is

$$S_{\rho SW} = -\sum_{i=1}^{4} \rho_{SW} n_H(i) U^*(i) \sigma_{NSW}(i) + \rho_{SW} n_H(3) U^*(3) \sigma_{NSW}(3) + \rho_{PUI} n_H(3) U^*(3) \sigma_{NPUI}(3) + \rho_{HSI} n_H(3) U^*(3) \sigma_{NHSI}(3), \quad (15)$$

for the PUIs it is

$$S_{\rho PUI} = \sum_{i=1}^{4} \rho_{SW} n_H(i) U^*(i) \sigma_{NSW}(i) + \rho_{HSI} n_H(i) U^*(i) \sigma_{NHSI}(i) - \rho_{SW} n_H(3) U^*(3) \sigma_{NSW}(3) - \rho_{PUI} n_H(3) U^*(3) \sigma_{NPUI}(3) - \rho_{HSI} n_H(3) U^*(3) \sigma_{NHSI}(3) \quad (16)$$

and for the heliosheath-created ions it is

$$S_{\rho HSI} = -\sum_{i=1}^{4} \rho_{HSI} n_H(i) U^*(i) \sigma_{NHSI}(i). \quad (17)$$

The source terms in density of the neutral populations *i*=1,2,4 and population 3 are:

$$S_{\rho H}(i \neq 3) = -\rho_{SW} n_H(i) U^*(i) \sigma_{NSW}(i) - \rho_{PUI} n_H(i) U^*(i) \sigma_{NPUI}(i) - \rho_{HSI} n_H(i) U^*(i) \sigma_{NHSI}(i) \quad (18)$$

$$S_{\rho H}(i = 3) = \sum_{i=1}^{4} \rho_{SW} n_H(i) U^*(i) \sigma_{NSW}(i) + \sum_{i=1}^{4} \rho_{PUI} n_H(i) U^*(i) \sigma_{NPUI}(i) + \sum_{i=1}^{4} \rho_{HSI} n_H(i) U^*(i) \sigma_{NHSI}(i) - \rho_{SW} n_H(3) U^*(3) \sigma_{NSW}(3) - \rho_{PUI} n_H(3) U^*(3) \sigma_{NPUI}(3) - \rho_{HSI} n_H(3) U^*(3) \sigma_{NHSI}(3). \quad (19)$$

In Region 3, in the supersonic solar wind, the momentum source term for the solar wind protons is

$$S_{M_{SW}} = -\sum_{i=1}^{4} \rho_{SW} n_H(i) {U_M}^*(i) \sigma_{SW}(i)\, U_{SW} + \rho_{SW} n_H(3) {U_M}^*(3) \sigma_{SW}(3) U_H(3) + \rho_{PUI} n_H(3) {U_M}^*(3) \sigma_{PUI}(3) U_H(3) + \rho_{HSI} n_H(3) {U_M}^*(3) \sigma_{HSI}(3) U_H(3), \quad (20)$$

for the PUIs it is

$$S_{M_{PUI}} = \sum_{i=1}^{4} \rho_{PUI} n_H(i) U_M{}^*(i) \sigma_{SW}(i) \left(U_H\,(i) - U_{PUI}\right) + \sum_{i=1}^{4} \rho_{SW} n_H(i) U_M{}^*(i) \sigma_{SW}(i)\, U_H(i) + \sum_{i=1}^{4} \rho_{HSI} n_H(i) U_M{}^*(i) \sigma_{HSI}(i)\, U_H(i) - \rho_{SW} n_H(3) U_M{}^*(3) \sigma_{SW}(3) U_H(3) - \rho_{PUI} n_H(3) U_M{}^*(3) \sigma_{PUI}(3) U_H(3) - \rho_{HSI} n_H(3) U_M{}^*(3) \sigma_{HSI}(3) U_H(3), \quad (21)$$

and for the heliosheath-created ions it is

$$S_{M_{HSI}} = -\sum_{i=1}^{4} \rho_{HSI} n_H(i) U_M{}^*(i) \sigma_{HSI}(i)\, U_{HSI}. \quad (22)$$

The momentum source terms of the neutral populations $i$=1, 2, 4 and population 3 are:

$$S_{M_H}(i \neq 3) = -\rho_{SW} n_H(i) U_M{}^*(i) \sigma_{SW}(i) U_H(i) - \rho_{PUI} n_H(i) U_M{}^*(i) \sigma_{PUI}(i) U_H(i) - \rho_{HSI} n_H(i) U_M{}^*(i) \sigma_{HSI}(i) U_H(i) \quad (23)$$

$$S_{M_H}(i = 3) = \sum_{i=1}^{4} \rho_{SW} n_H(i) U_M{}^*(i) \sigma_{SW}(i)\, U_{SW} + \sum_{i=1}^{4} \rho_{PUI} n_H(i) U_M{}^*(i) \sigma_{PUI}(i)\, U_{PUI} + \sum_{i=1}^{4} \rho_{HSI} n_H(i) U_M{}^*(i) \sigma_{HSI}(i)\, U_{HSI} - \rho_{SW} n_H(3) U_M{}^*(3) \sigma U_H(3) - \rho_{PUI} n_H(3) U_M{}^*(3) \sigma U_H(3) - \rho_{HSI} n_H(3) U_M{}^*(3) \sigma U_H(3). \quad (24)$$

In Region 3, in the supersonic solar wind, the energy source term for the solar wind protons is

$$S_{\varepsilon_{SW}} = -\sum_{i=1}^{4} [0.5 \rho_{SW} n_H(i) U_M{}^*(i) \sigma_{SW}(i) U_{SW}^2 + \rho_{SW} n_H(i) U^*(i) \sigma_{SW}(i) U_{thSW}] + 0.5 \rho_{SW} n_H(3) U_M{}^*(3) \sigma_{SW}(3) U_H^2(3) + \rho_{SW} n_H(3) U^*(3) \sigma_{SW}(3) U_{th}(3) + 0.5 \rho_{PUI} n_H(3) U_M{}^*(3) \sigma_{PUI}(3) U_H^2(3) + \rho_{PUI} n_H(3) U^*(3) \sigma_{PUI}(3) U_{th}(3) + 0.5 \rho_{HSI} n_H(3) U_M{}^*(3) \sigma_{HSI}(3) U_H^2(3) + \rho_{HSI} n_H(3) U^*(3) \sigma_{HSI}(3) U_{th}(3), \quad (25)$$

for the PUIs it is

$$S_{\varepsilon_{PUI}} = \sum_{i=1}^{4} [0.5 \rho_{PUI} n_H(i) U_M{}^*(i) \sigma_{PUI}(i) (U_H^2(i) - U_{PUI}^2) + \rho_{PUI} n_H(i) U^*(i) \sigma_{PUI}(i) (U_{th}(i) - U_{thPUI})] - 0.5 \rho_{PUI} n_H(3) \sigma_{PUI}(3) U_M{}^*(3) (U_H^2(3) - U_{PUI}^2) - \rho_{PUI} n_H(3) U^*(3) \sigma_{PUI}(3) (U_{th}(3) - U_{thPUI}) + \sum_{i=1}^{4} [0.5 \rho_{SW} n_H(i) U_M{}^*(i) \sigma_{SW}(i) U_H^2(i) + \rho_{SW} n_H(i) U^*(i) \sigma_{SW}(i) U_{th}(i)] - 0.5 \rho_{SW} n_H(3) U_M{}^*(3) \sigma_{SW}(3) U_H^2(3) - \rho_{SW} n_H(3) U^*(3) \sigma_{SW}(3) U_{th}(3) + \sum_{i=1}^{4} [0.5 \rho_{HSI} n_H(i) U_M{}^*(i) \sigma_{HSI}(i) U_H^2(i) +$$

$$\rho_{HSI} n_H(i) U^*(i) \sigma_{HSI}(i) U_{th}(i)] - 0.5\rho_{HSI} n_H(3) U_M{}^*(3) \sigma_{HSI}(3) U_H^2(3) -$$

$$\rho_{HSI} n_H(3) U^*(3) \sigma_{HSI}(3) U_{th}(3), \quad (26)$$

and for the heliosheath-created ions it is

$$S_{\varepsilon_{HSI}} = -\sum_{i=1}^{4} [0.5\rho_{HSI} n_H(i) U_M{}^*(i) \sigma_{HSI}(i) U_{HSI}^2 + \rho_{HSI} n_H(i) U^*(i) \sigma_{HSI}(i) U_{thHSI}]. \quad (27)$$

The energy source terms of the neutral populations i=1, 2, 4 and population 3 are:

$$S_{\varepsilon_H}(i \neq 3) = -0.5\rho_{SW} n_H(i) U_M{}^*(i) \sigma_{SW}(i) U_H^2(i) - \rho_{SW} n_H(i) U^*(i) \sigma_{SW}(i) U_{th}(i) -$$

$$0.5\rho_{PUI} n_H(i) U_M{}^*(i) \sigma_{PUI}(i) U_H^2(i) - \rho_{PUI} n_H(i) U^*(i) \sigma_{PUI}(i) U_{th}(i) -$$

$$0.5\rho_{HSI} n_H(i) U_M{}^*(i) \sigma_{HSI}(i) U_H^2(i) - \rho_{HSI} n_H(i) U^*(i) \sigma_{HSI}(i) U_{th}(i) \quad (28)$$

$$S_{\varepsilon_H}(i = 3) = -0.5\rho_{SW} n_H(3) U_M{}^*(3) \sigma_{SW}(3) U_H^2(3) - \rho_{SW} n_H(3) U^*(3) \sigma_{SW}(i) U_{th}(3) -$$

$$0.5\rho_{PUI} n_H(3) U_M{}^*(3) \sigma_{PUI}(3) U_H^2(3) - \rho_{PUI} n_H(3) U^*(3) \sigma_{PUI}(3) U_{th}(3) -$$

$$0.5\rho_{HSI} n_H(3) U_M{}^*(3) \sigma_{HSI}(3) U_H^2(3) - \rho_{HSI} n_H(3) U^*(3) \sigma_{HSI}(3) U_{th}(3) +$$

$$\sum_{i=1}^{4} [\rho_{SW} n_H(i) U_M{}^*(i) \sigma_{SW}(i) U_{SW}^2 + \rho_{SW} n_H(i) U^*(i) \sigma_{SW}(i) U_{thSW}] +$$

$$\sum_{i=1}^{4} \left[\rho_{PUI} n_H(i) U_M{}^*(i) \sigma_{PUI}(i) U_{PUI}^2 + \rho_{PUI} n_H(i) U^*(i) \sigma_{PUI}(i) U_{thPUI}\right] +$$

$$\sum_{i=1}^{4} \left[\rho_{HSI} n_H(i) U_M{}^*(i) \sigma_{HSI}(i) U_{HSI}^2 + \rho_{HSI} n_H(i) U^*(i) \sigma_{HSI}(i) U_{thHSI}\right]. \quad (29)$$

**A1.3 MHD Model: Region 2 (Heliosheath) Source Terms**

In Region 2, in the heliosheath, the density source term for the solar wind protons is

$$S_{\rho SW} = -\sum_{i=1}^{4} \rho_{SW} n_H(i) U^*(i) \sigma_{NSW}(i), \quad (30)$$

for the PUIs it is

$$S_{\rho PUI} = -\sum_{i=1}^{4} \rho_{PUI} n_H(i) U^*(i) \sigma_{NPUI}(i), \quad (31)$$

and for the heliosheath-created ions it is

$$S_{\rho HSI} = \sum_{i=1}^{4} [\rho_{SW} n_H(i) U^*(i) \sigma_{NSW}(i) + \rho_{PUI} n_H(i) U^*(i) \sigma_{NPUI}(i)]. \quad (32)$$

The density source terms of the neutral populations *i*=1, 3, 4 and population 2 are:

$$S_{\rho H}(i \neq 2) = -\rho_{SW} n_H(i) U^*(i) \sigma_{NSW}(i) - \rho_{PUI} n_H(i) U^*(i) \sigma_{NPUI}(i) - \rho_{HSI} n_H(i) U^*(i) \sigma_{NHSI}(i)$$

(33)

$$S_{\rho H}(i = 2) = \sum_{i=1}^{4} \rho_{SW} n_H(i) U^*(i) \sigma_{NSW}(i) + \sum_{i=1}^{4} \rho_{PUI} n_H(i) U^*(i) \sigma_{NPUI}(i) + \sum_{i=1}^{4} \rho_{HSI} n_H(i) U^*(i) \sigma_{NHSI}(i) - \rho_{SW} n_H(2) U^*(2) \sigma_{NSW}(2) - \rho_{PUI} n_H(2) U^*(2) \sigma_{NPUI}(2) - \rho_{HSI} n_H(2) U^*(2) \sigma_{NHSI}(2). \quad (34)$$

In Region 2, in the heliosheath, the momentum source term for the solar wind protons is

$$S_{M_{SW}} = -\sum_{i=1}^{4} \rho_{SW} n_H(i) {U_M}^*(i) \sigma_{SW}(i) U_{SW}, \quad (35)$$

for the PUIs it is

$$S_{M_{PUI}} = -\sum_{i=1}^{4} \rho_{PUI} n_H(i) {U_M}^*(i) \sigma_{PUI}(i) U_{PUI}, \quad (36)$$

and for the heliosheath-created ions it is

$$S_{M_{HSI}} = \sum_{i=1}^{4} \rho_{HSI} n_H(i) {U_M}^*(i) \sigma_{HSI}(i) \left(U_H(i) - U_{HSI}\right) + \sum_{i=1}^{4} \rho_{SW} n_H(i) {U_M}^*(i) \sigma_{SW}(i) U_H(i) + \sum_{i=1}^{4} \rho_{PUI} n_H(i) {U_M}^*(i) \sigma_{PUI}(i) U_H(i). \quad (37)$$

The momentum source terms of the neutral populations *i*=1, 3, 4 and population 2 are:

$$S_{M_H}(i \neq 2) = -\rho_{SW} n_H(i) {U_M}^*(i) \sigma U_H(i) - \rho_{PUI} n_H(i) {U_M}^*(i) \sigma U_H(i) - \rho_{HSI} n_H(i) {U_M}^*(i) \sigma U_H(i) \quad (38)$$

$$S_{M_H}(i = 2) = \sum_{i=1}^{4} \rho_{SW} n_H(i) {U_M}^*(i) \sigma_{SW}(i) U_{SW} + \sum_{i=1}^{4} \rho_{PUI} n_H(i) {U_M}^*(i) \sigma_{PUI}(i) U_{PUI} + \sum_{i=1}^{4} \rho_{HSI} n_H(i) {U_M}^*(i) \sigma_{HSI}(i) U_{HSI} - \rho_{SW} n_H(2) {U_M}^*(2) \sigma_{SW}(2) U_H(2) - \rho_{PUI} n_H(2) {U_M}^*(2) \sigma_{PUI}(2) U_H(2) - \rho_{HSI} n_H(2) {U_M}^*(2) \sigma_{HSI}(2) U_H(2). \quad (39)$$

In Region 2, in the heliosheath, the energy source term for the solar wind protons is

$$S_{\mathcal{E}_{SW}} = -\sum_{i=1}^{4} \left[0.5 \rho_{SW} n_H(i) {U_M}^*(i) \sigma_{SW}(i) U_{SW}^2 + \rho_{SW} n_H(i) U^*(i) \sigma_{SW}(i)\, U_{thSW}\right], \quad (40)$$

for the PUIs it is

$$S_{\mathcal{E}_{PUI}} = -\sum_{i=1}^{4} \left[0.5 \rho_{PUI} n_H(i) {U_M}^*(i) \sigma_{PUI}(i) U_{PUI}^2 + \rho_{PUI} n_H(i) U^*(i) \sigma_{PUI}(i)\, U_{thPUI}\right], \quad (41)$$

and for the heliosheath-created ions it is

$$S_{\mathcal{E}_{HSI}} = \sum_{i=1}^{4}[0.5\rho_{HSI} n_H(i) U_M{}^*(i)\sigma_{HSI}(i)(U_H^2(i) - U_{HSI}^2) + \rho_{HSI} n_H(i) U^*(i)\sigma_{HSI}(i)(U_{th}(i) - U_{thHSI})] + \sum_{i=1}^{4}[0.5\rho_{SW} n_H(i) U_M{}^*(i)\sigma_{SW}(i) U_H^2(i) + \rho_{SW} n_H(i) U^*(i)\sigma_{SW}(i) U_{th}(i)] + \sum_{i=1}^{4}[0.5\rho_{PUI} n_H(i) U_M{}^*(i)\sigma_{PUI}(i) U_H^2(i) + \rho_{PUI} n_H(i) U^*(i)\sigma_{PUI}(i) U_{th}(i)]. \quad (42)$$

The energy source terms of the neutral populations *i*=1, 3, 4 and population 2 are:

$$S_{\mathcal{E}_H}(i \neq 2) = -0.5\rho_{SW} n_H(i) U_M{}^*(i)\sigma_{SW}(i) U_H^2(i) - \rho_{SW} n_H(i) U^*(i)\sigma_{SW}(i) U_{th}(i) - 0.5\rho_{PUI} n_H(i) U_M{}^*(i)\sigma_{PUI}(i) U_H^2(i) - \rho_{PUI} n_H(i) U^*(i)\sigma_{PUI}(i) U_{th}(i) - 0.5\rho_{HSI} n_H(i) U_M{}^*(i)\sigma_{HSI}(i) U_H^2(i) - \rho_{HSI} n_H(i) U^*(i)\sigma_{HSI}(i) U_{th}(i) \quad (43)$$

$$S_{\mathcal{E}_H}(i = 2) = \sum_{i=1}^{4}[\rho_{SW} n_H(i) U_M{}^*(i)\sigma_{SW}(i) U_{SW}^2 + \rho_{SW} n_H(i) U^*(i)\sigma_{SW}(i) U_{th}(i)] + \sum_{i=1}^{4}[\rho_{PUI} n_H(i) U_M{}^*(i)\sigma_{PUI}(i) U_{PUI}^2 + \rho_{PUI} n_H(i) U^*(i)\sigma_{PUI}(i) U_{thPUI}] + \sum_{i=1}^{4}[\rho_{HSI} n_H(i) U_M{}^*(i)\sigma_{HSI}(i) U_{HSI}^2 + \rho_{HSI} n_H(i) U^*(i)\sigma_{HSI}(i) U_{thHSI}] - 0.5\rho_{SW} n_H(2) U_M{}^*(2)\sigma_{SW}(2) U_H^2(2) - \rho_{SW} n_H(2) U^*(2)\sigma_{SW}(2) U_{th}(2) - 0.5\rho_{PUI} n_H(2) U_M{}^*(2)\sigma_{PUI}(2) U_H^2(2) - \rho_{PUI} n_H(2) U^*(2)\sigma_{PUI}(2) U_{th}(2) - 0.5\rho_{HSI} n_H(2) U_M{}^*(2)\sigma_{HSI}(2) U_H^2(2) - \rho_{HSI} n_H(2) U^*(2)\sigma_{HSI}(2) U_{th}(2) \quad (44)$$

### A1.4 MHD Model: Region 1 (Outer Heliosheath) Source Terms

In Region 1, the outer heliosheath source terms are the same as in Region 2, except the index of “2” is replaced with the index of “1” in Equations 30-44.

## A2. ENA MODEL

We update our previous ENA model (M. Kornbleuth et al. 2023b). In this model, the plasma is partitioned at the termination shock into multiple ion species. The density, energy, and the form of the distributions for the separate ion species are evaluated by fitting to the hybrid simulation results of ion acceleration at the termination shock (J. Giacalone et al. 2021). In our previous work (M. Kornbleuth et al. 2023b), a single-ion plasma is partitioned into four separate ion populations: thermal solar wind ions and PUIs that are created in the supersonic solar wind and are either transmitted or reflected at the termination shock, with some of the reflected PUIs undergoing further acceleration via diffusive shock acceleration (B. Wang et al. 2023). Once inside the heliosheath, the ion species of varying energies are removed via charge exchange, which leads to a depletion of high energy ions at large distances into the heliosheath known as extinction. This model only considers ion acceleration at the termination shock. An example output from using this model can be found in M. Kornbleuth et al. (2023a), which demonstrates that acceleration at the termination shock alone is insufficient for reproducing the Belt morphology.

As an update to our previous work (M. Kornbleuth et al. 2023b), we are using a multi-ion MHD model. Therefore, we can more self-consistently model the PUI distributions at the termination shock. M. Kornbleuth et al. (2023b) partitioned the plasma into thermal solar wind ions and PUIs, and then partitioned the PUIs further into the different components. Here, we have separate thermal solar wind and PUI components at the termination shock, so we partition the PUI fluid at the termination shock into the three components: transmitted PUIs, reflected PUIs, and reflected PUIs accelerated via diffusive shock acceleration (B. Wang et al. 2023). The density fractions used to partition the PUI fluids from the supersonic solar wind are determined by the fractions used to fit hybrid modeling of PUIs at the termination shock (M. Kornbleuth et al. 2023b; J. Giacalone et al. 2021). In terms of the energy fraction, it was shown previously that the multi-ion model, when treating the electrons separately, overpredicts the thermal solar wind temperature due to the way energy is distributed across the termination shock (E. S. Bair et al 2025). As such, in the MHD model the thermal solar wind ions have too much energy while the PUIs from the supersonic solar wind are lacking in energy relative to Voyager observations (J.

D. Richardson et al. 2008). To account for this discrepancy, we therefore use the total plasma temperature to determine the energy of each species by using the previous ENA modeling method to fit the hybrid results (M. Kornbleuth et al. 2023b). We identify the energy fraction of each of the ion species in our ENA model relative to the plasma and calculate the temperature using the energy fraction and density of the population accordingly.

As an improvement to M. Kornbleuth et al. (2023b), we now consider variations in the upstream PUI speed at the termination shock. Previously, our results only considered slow solar wind upstream of the termination shock (308-333 km/s) based on hybrid simulations of locations either in the ecliptic plane or Voyager 2 observations (J. Giacalone et al. 2021). Recent updates to the hybrid modeling (J. Giacalone et al. 2025) considered the effects of different upstream PUI speeds in the Voyager 2 direction, sampling speeds of 320 km/s, 480 km/s, and a mixture case consisting of a partially filled PUI shell distribution with a radius varying from 320 km/s to 640 km/s. Considering our solar wind conditions include latitudinal variations with fast solar wind at the poles, we apply the mixture case at the poles to reflect a 22-year average over solar minimum and maximum conditions. While we do expect changes in the form of the downstream distribution function depending on the shock-normal angle of the termination shock, field-line meandering due to large-scale turbulence in the simulations can lead to variations in the angle between the magnetic field and normal direction at the shock, which can cause similar local geometries despite different shock-normal angles. Based on this assumption, we apply the mixture case for Voyager 2 from the updated hybrid simulations (J. Giacalone et al. 2025) at high latitudes ($|\theta| > 45°$). As we expect a smooth latitudinal transition in the evolution of the downstream PUI distribution, we apply an interpolation for the fits to the hybrid simulation (M.

Kornbleuth et al. 2023b) from the low-latitude region (J. Giacalone et al. 2021) to the high-latitude region (J. Giacalone et al. 2025).

For a given fitting parameter, Q, we need to interpolate from the low-latitude region to the high-latitude region. For the low-latitude region of slow solar wind, we have $Q_{21}$ and for the high-latitude region of fast solar wind, we have $Q_{25}$. We initiate the transition to fast wind at ($|\theta| > 45°$), where $\theta$ is the angle from the ecliptic plane. At $|\theta| = 90°$, we have $Q_{25}$, and at $|\theta| < 45°$ we have $Q_{21}$, derived from previous works (M. Kornbleuth et al. 2023b). To interpolate, we do a linear interpolation such that

$$\Delta Q = Q_{25} - Q_{21} \tag{45}$$

$$\Delta Q' = \Delta Q \frac{|\theta - \theta_{min}|}{|\theta_{max} - \theta_{min}|} = Q - Q_{21} \tag{46}$$

$$Q = \Delta Q \frac{|\theta - \theta_{min}|}{|\theta_{max} - \theta_{min}|} + Q_{21} \tag{47}$$

Here, $\theta_{min}$ and $\theta_{max}$ are 45° and 90°, respectively. By using this technique, we can obtain a smooth transition from the low to the high latitudes for parameters such as the reflected PUI density and energy fraction and the slope of the power law tail.

We propagate the different PUI populations into the heliosheath along the PUI flow streamlines, where they undergo charge exchange and extinction processes. The MHD solution provides an extinction profile for the PUI species originating from the supersonic solar wind. While this extinction profile is not necessarily applicable to all energies (because charge exchange is an energy-dependent process) the solution approximately replicates the extinction profile for a species with ~1 keV. Considering the contributing energy range for the PUIs at the termination shock is on the order of ~1-4 keV, we use the extinction from the MHD solution. This is because, while not equal, the cooling lengths for this energy range vary by 10-20 au (M. Kornbleuth et al. 2023a; N. A. Schwadron et al. 2014) while the overall size of the heliotail is on the order of hundreds of au. Therefore, we do not expect this to significantly influence our results.

As a further update to M. Kornbleuth et al. (2023b), we model the heliosheath-created ions present in our new MHD model. We treat these ions with a Maxwellian distribution and extract the conditions from the MHD solution to model ENAs from this species. I.Baliukin et al. (2020) showed that the distribution of heliosheath-created ions is not Maxwellian. Understanding the effect of the distribution for this population is left to future work. We do not need to apply additional extinction terms to the heliosheath-created ion population in our ENA model, as these ions are constantly replenished in the heliosheath through charge exchange between the overall heliosheath plasma and the background neutral populations.

In Fig. 5, we show a comparison between our previous single-ion MHD ENA modeling results (M. Kornbleuth et al. 2023a) with our new, updated results both with and without the inclusion of the reconnection process. For our ENA maps in Fig. 5 (as well as Fig. 2), we employ a boxcar smoothing over 3 grid cells to eliminate noise from the kinetic neutral solution. We find that the multi-ion MHD model with the updated hybrid results being used for the high latitudes gives notable improvement compared to the single-ion results. While we will do a detailed comparison and analysis of the changes between the models from the perspective of MHD and ENA modeling in a future work, we find that the self-consistent treatment of PUIs in the multi-ion MHD model allows for a more realistic approximation for modeling the ion distribution at the termination shock and in the heliosheath. Likewise, using the updated hybrid results at high latitudes provides a more physical representation of the solar wind upstream of the termination shock at high latitudes, which also enhances the comparison. Interestingly, while reconnection has the largest impact on the ENA maps at the INCA energies, we do find non-negligible contributions at lower energies in low latitudes, which also contributes to a better comparison with the observations. This may indicate that IBEX-Hi is also able to see evidence of reconnection in the heliosheath in addition to INCA.

Within our MHD solution, we identify the region where $mV_A^2$ is greater than the local thermal energy of the plasma, which is an important factor for reconnection to yield ion acceleration in the heliospheric tail. Within the region where $mV_A^2$ is high (e.g. Fig. 3), we adapt the fitted distributions from the *kglobal* model and apply them in our ENA model. While the low-β region correlates with a region with optimal conditions for reconnection, because we do not explicitly model the current sheet within our MHD we use $mV_A^2/E_{th} > 1$ as our criteria since this reflects where there is sufficient energy for particle energization via reconnection. Within the reconnection region, the plasma is predominantly composed of heliosheath-created ions because most of the ions originating from the supersonic solar wind at the relevant energies have been removed via charge exchange (Fig. 6). Therefore, within this region we can approximate the plasma as having two components: the unaccelerated and accelerated components of the heliosheath-created ions. We assume a constant reconnection process within this region, with the power-law index of the distribution becoming stable within 2 au of entering the reconnection region.

Since we are considering ENAs created from neutral-ion interactions, in Fig. 4 we show the spectra of the ions resulting from reconnection. At the energies of interest for modeling the ENAs involved in the emergence and disappearance of the Belt in the low-latitude tail (~4-20 keV), the power law for the ion reconnection spectrum has an approximate index of –3.60. In comparison, the power-law index best reflecting the ENA observations from 2009-2012 in the energy range 4-20 keV is –4.11 ± 0.05. It is important to note, however, that here we simulate an average distribution that corresponds to a time period just before the manifestation of the solar minimum in the heliosheath. As previously explained (K. Dialynas et al. 2017), the slopes of the ENA spectra sampled from the Belt decrease over time following the corresponding ENA intensity decrease and, furthermore, as we approach solar minimum the spectra of the Belt and basin (the minima of ENA emissions in INCA observations) become almost indistinguishable. Typically, the spectral index of ENA spectra in the basin varies between -3.4 and -4.0 (K. Dialynas et al. 2013a).

To generate the fit to the reconnection spectrum, we fit the *kglobal* result (Zhiyu Yin et al. 2024b) with a guide field of $B_g/B_0 = 0.75$. As shown in Fig. 7, a strong guide field is required to replicate the energy dependence of the Belt in the low-latitude heliotail. For weak guide fields (i.e. $B_g/B_0 = 0.25$), the low-latitude heliotail has a harder spectrum and strengthens rather than weakens in the 18.0 keV energy band. For the heliosheath-created ions, we assume, based on our fit, that 1.4% of the ions are accelerated via reconnection. To properly model this accelerated component, we fit two kappa distributions. For the first kappa distribution, we assign 96% of the accelerated ions by density to have a kappa value of 8.60, with a temperature 53 times the ambient heliosheath-created ion temperature. For the second kappa distribution, we assign 4% of the accelerated ions by density to have a kappa value of 8.15, with a temperature 177 times the ambient heliosheath-created ion temperature. We note that these are empirical fits to the reconnection spectrum for the purposes of implementation into the ENA model and we do not derive any physical interpretation of the values required to replicate the fit. We apply the fits within the low-$\beta$ region of the heliospheric tail within our ENA model. We can then use this accelerated ion distribution to calculate the ENA flux.

In our ENA model, we initially assume the heliosheath-created ions to have a Maxwellian distribution when entering the low-β region. Based on average values within the low-β region of our MHD simulation, the timescale for reconnection to develop a Maxwellian proton distribution into a stable power law is $t_m \sim 0.2$ years, which is comparable to the timescale for plasma flow across one 2 au grid cell in the low-β region of our ENA simulation. Considering the low-β region has a radial size >> 2 au, we can then apply our modified distributions to consider the transformation of the spectra resulting from magnetic reconnection. However, there is a need to further explore this assumption. As shown in previous INCA observations (K. Dialynas et al. 2015), there is an ENA flux gradient between the Belt and the basins, which is not replicated within our ENA modeling. This may be due to the assumption that reconnection effects on ion acceleration, including the guide field, are the same for the entire reconnection region. Alternatively, this gradient may actually be steeper than noted based on limitations from the angular resolution of the measurement. In reality, there are likely different conditions within the reconnection region which would contribute to ion acceleration differently and could contribute to the observed gradient.

**Acknowledgements**

We would like to thank E. Bair for discussions on the development of the MHD model and J. Miller for discussions on implications. The authors are supported by NASA (Grant No. 18-DRIVE18_2-0029), Our Heliospheric Shield, 80NSSC22M0164. For more information about this center, please visit https://shielddrivecenter.com. KD also acknowledges support at JHU/APL by NASA under contracts NAS597271, NNX07AAJ69G, and NNN06AA01C and by subcontract at the Center of Space Research and Technology. JFD, MS, and ZY also acknowledge support from NSF grant PHY2109083, as well as NASA grants 80NSSC22M0164 and 80NSSC20K1277.

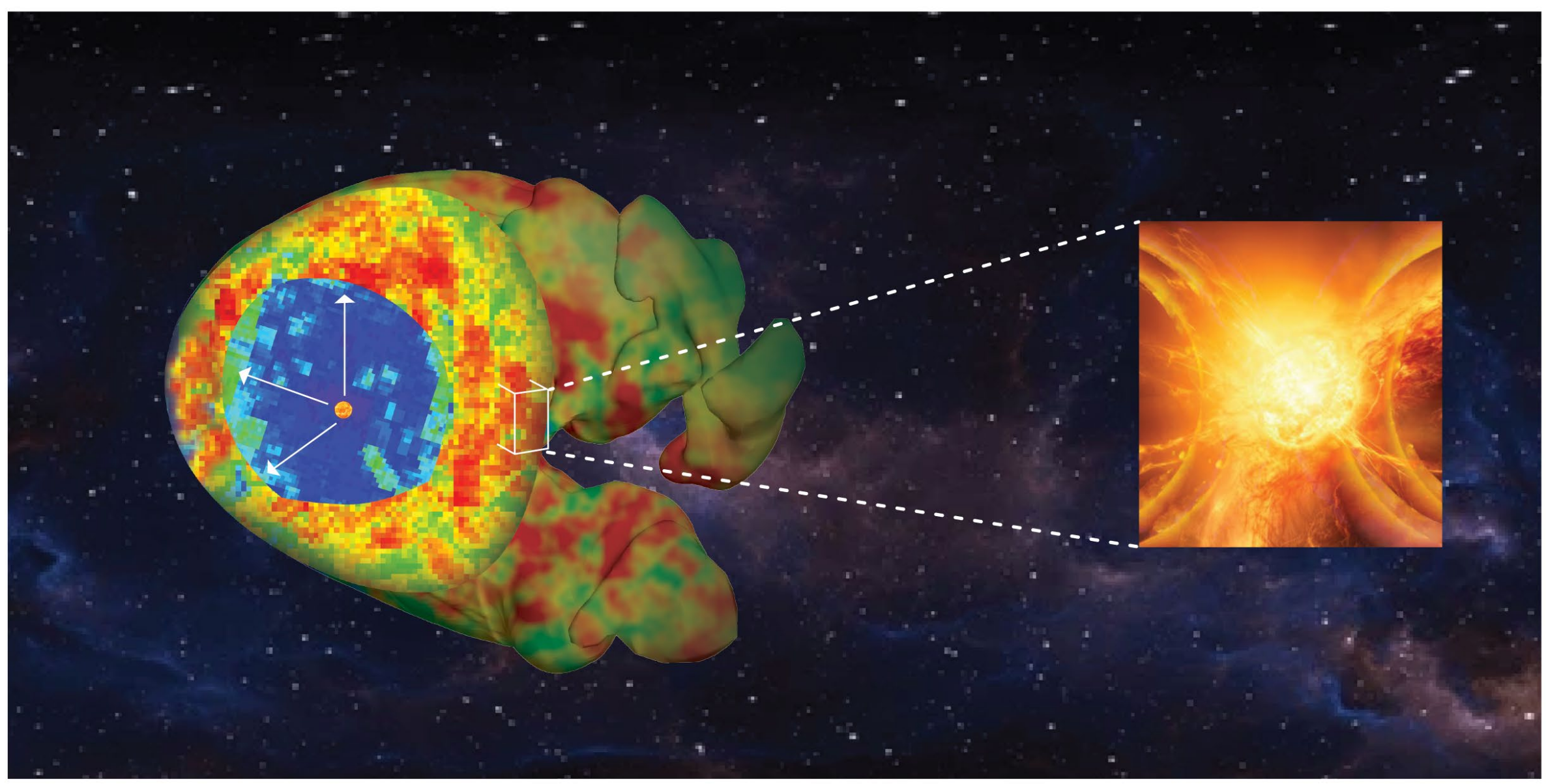

**Fig. 1:** Magnetic reconnection in the split tail heliosphere creates the Buckle. We show a cartoon depicting reconnection taking place deep in the low-latitude heliotail. The shading across the heliosphere reflects an illustration of the INCA ENA Belt. The zoomed box indicates the region in the tail of interest, where the plasma is low-β and reconnection is occurring. The reconnection process, illustrated within the magnified box, is energizing ions and creating the ENAs within the Buckle of the Belt.

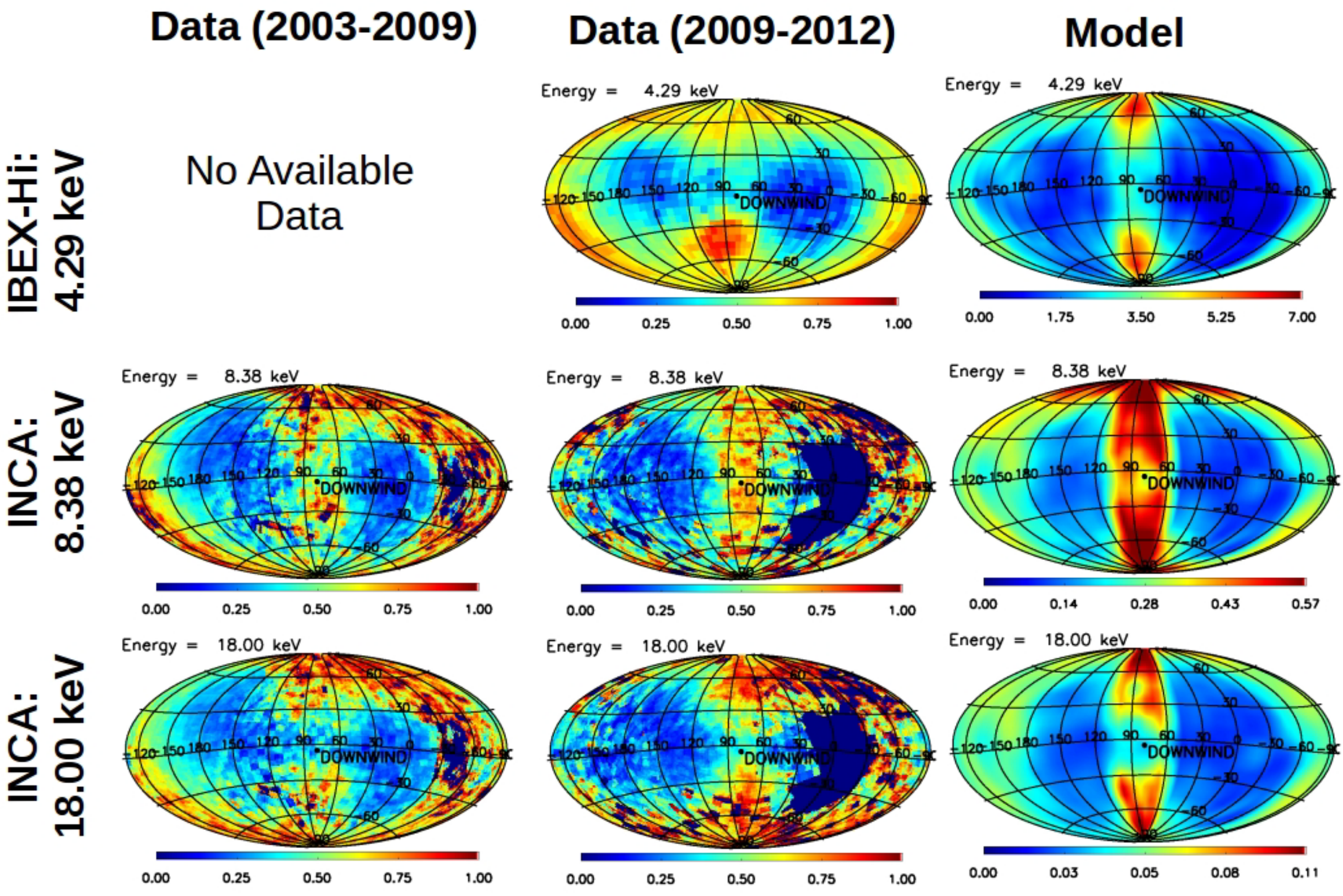

**Fig. 2:** Magnetic reconnection in the heliotail replicates the Belt morphology seen in ENA observations. This is the first work to replicate the energy-dependence of the Belt with modeling. We present a comparison of ENA observations and model results for the 4.29, 8.38, and 18.00 keV energy bands centered on the heliotail. The 4.29 keV energy band corresponds to IBEX-Hi observations. The 8.38 and 18.00 keV energy bands correspond to INCA observations. Left column: ENA observations corresponding to an average over the years 2003-2009. We note only INCA data is included since the IBEX satellite was not launched until 2009. Middle column: ENA observations corresponding to an average over the years 2009-2012. Right column: ENA model results including reconnection in the heliotail. For all cases, the Buckle appears at 8.38 keV, and otherwise disappears at lower and higher energies. All maps derived from data are normalized relative to the maximum flux within each individual map to provide a qualitative comparison of the morphology. The units of the simulation results are ENAs/($cm^2$ s sr keV).

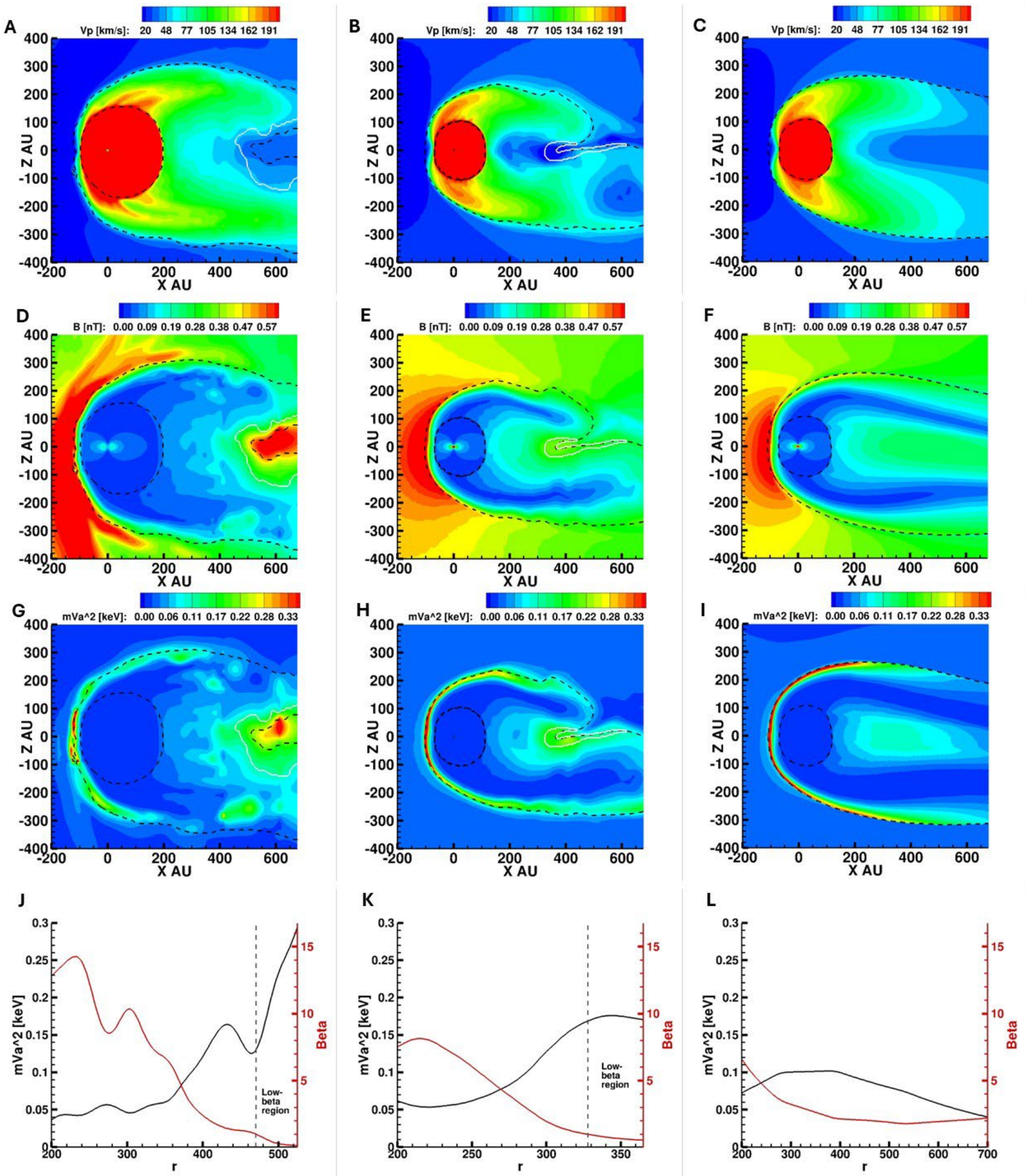

**Fig. 3:** The low-β region in the heliotail is a feature of the split tail heliosphere. We show a comparison of the results of three simulation models: the plasma speed (A-C), magnetic field strength (D-F), and magnetic energy per particle, $mV_A^2$ (G-I) in a meridional slice of the heliosphere. In panels A-I, the white line contour (if present) corresponds to β=1 in the heliotail, and the inner and outer dashed black lines correspond to the termination shock and heliopause, respectively. In panels J-L, we present a comparison of the $mV_A^2$ and plasma-β along a cut through the low-latitude heliotail. The black line in panels J-L corresponds to $mV_A^2$ and the red line corresponds to the plasma-β. A black-dashed line is included in J-L to show the onset of the low-β region in each model if a low-β region is present. Panels A/D/G/J correspond to the multi-ion MHD model of the split-tail heliosphere highlighted in this work. Panels B/E/H/K correspond to a single-ion MHD model of the split tail heliosphere (M. Kornbleuth et al. 2021a). Panels C/F/I/L correspond to a single-ion MHD model of a long-tail heliosphere without helium ions included (V. V. Izmodenov & D. B. Alexashov 2020). The single-ion MHD models produce different termination shock and heliopause locations than the multi-ion model due to different interstellar conditions and pressure differences in the heliosheath. As shown, due to a lack of pile-up of the magnetic field in the low-latitude heliotail in the long-tail heliosphere, a region of low-β (β < 1) is not present.

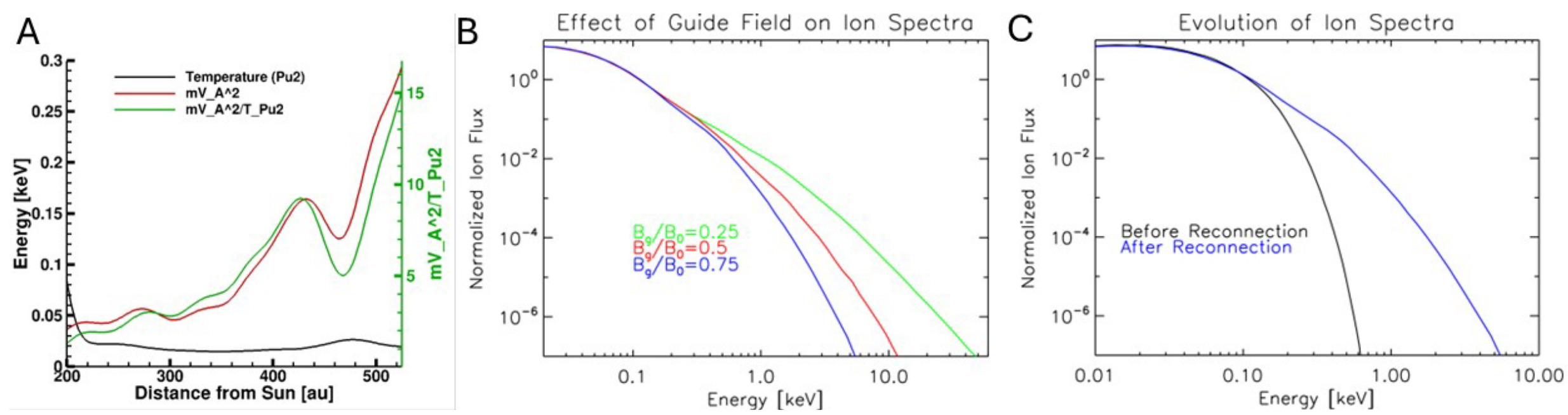

**Fig. 4:** Reconnection in the heliotail can accelerate cold (~eV) ions to ~keV energies. Panel A shows the comparison along a 1D cut through the low latitude heliotail of the thermal energy (black line) of heliosheath-created ions in the heliotail relative to the available magnetic energy ($mV_A^2$; red line). A high value of $mV_A^2$ indicates a large amount of available magnetic energy during reconnection (T. D. Phan et al. 2013, 2014). There is a higher abundance of magnetic energy relative to the thermal energy (green line), indicating the potential for substantial energization of the ions during reconnection. Panel B shows the effect of a strong guide field ($B_g/B_0$) on the accelerated ion spectrum. Panel C shows the initial and final form of the ion spectrum in a *kglobal* simulation for $B_g/B_0 = 0.75$, which is used in this work, showing the development of a power-law tail.

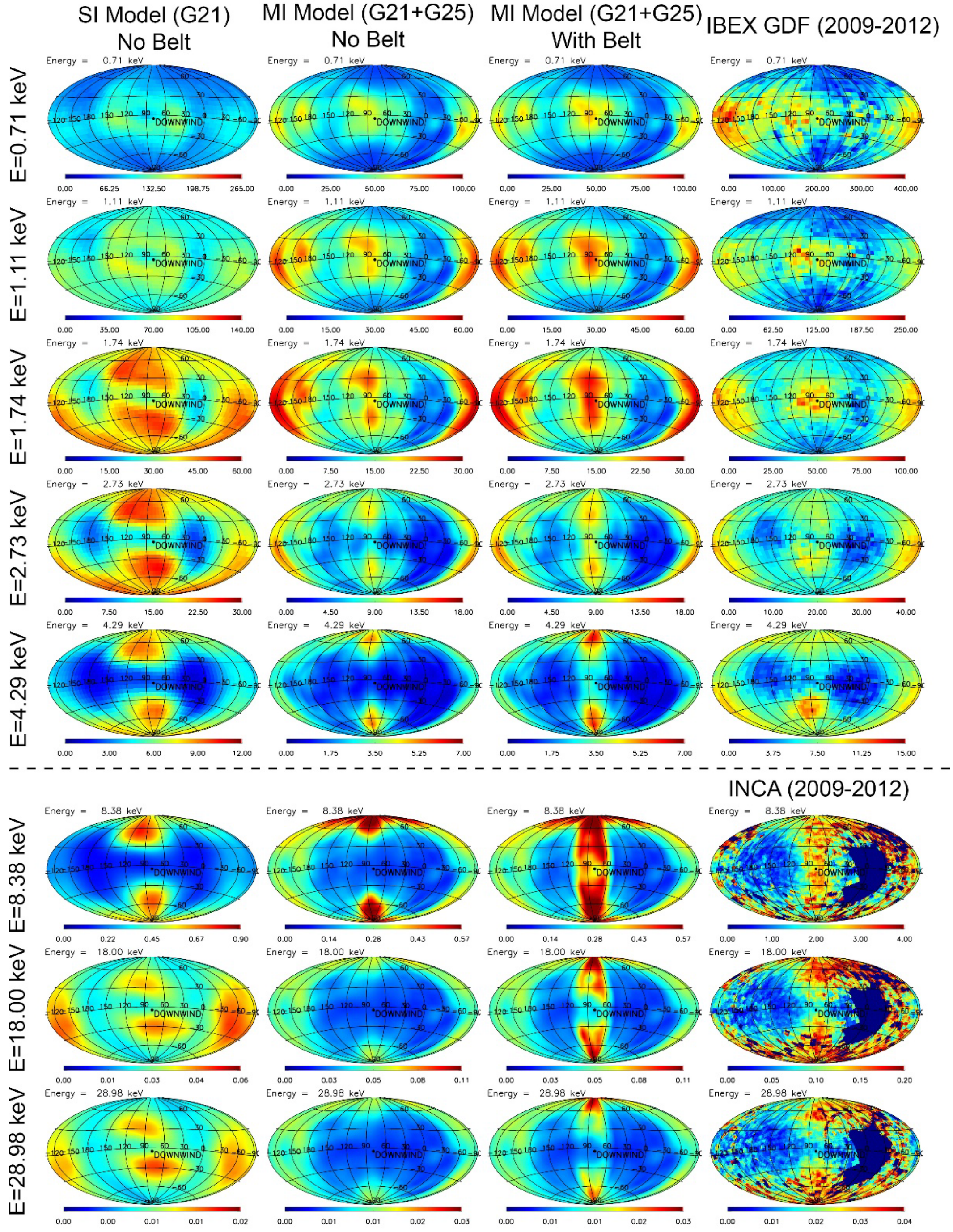

**Fig. 5**: The Multi-ion MHD model provides an improved comparison with IBEX and INCA observations. All ENA maps are tail-centered with units of ENAs/(cm$^2$ s sr keV). Left: ENA results from a single-ion MHD model (M. Kornbleuth et al. 2023a) at IBEX-Hi energies. These model results do not consider reconnection within the heliosheath, and do not consider solar wind speed variations upstream of the termination shock. Middle-left: ENA results from the multi-ion MHD model described in this work, using solar wind speed variations upstream of the termination shock. These maps do not include reconnection within the heliosheath. Middle-right: same as the middle-left column, but with the reconnection process included. The results suggest that, while not strong, reconnection could be affecting the ENA flux signal in the low-latitude

tail at even the lowest IBEX-Hi energies. At INCA energies, the results suggest that including reconnection in the heliotail is critical for replicating the energy dependence and morphology of the Belt. Right: ENA observations from IBEX-Hi and INCA averaged over the years 2009-2012. The IBEX data is the ribbon-separated data from the 18th IBEX data release (D. J. McComas et al. 2024).

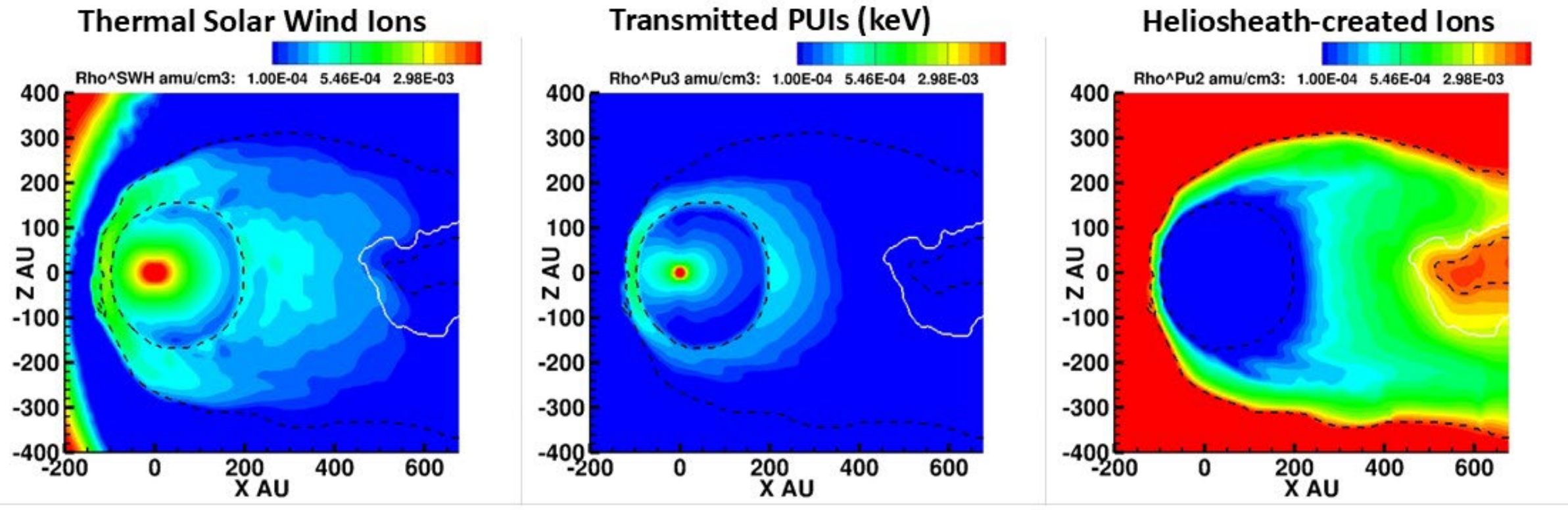

**Fig. 6:** The heliosheath-created ion population is the dominant population by density in the deep heliotail. A comparison of the densities of the thermal solar wind ions (left), transmitted PUIs (middle) and heliosheath-created ions (right) in the multi-ion MHD model shown in the meridional plane. At ~400 au from the Sun in the heliotail, the heliosheath-created ions are approximately 75% of the total plasma density.

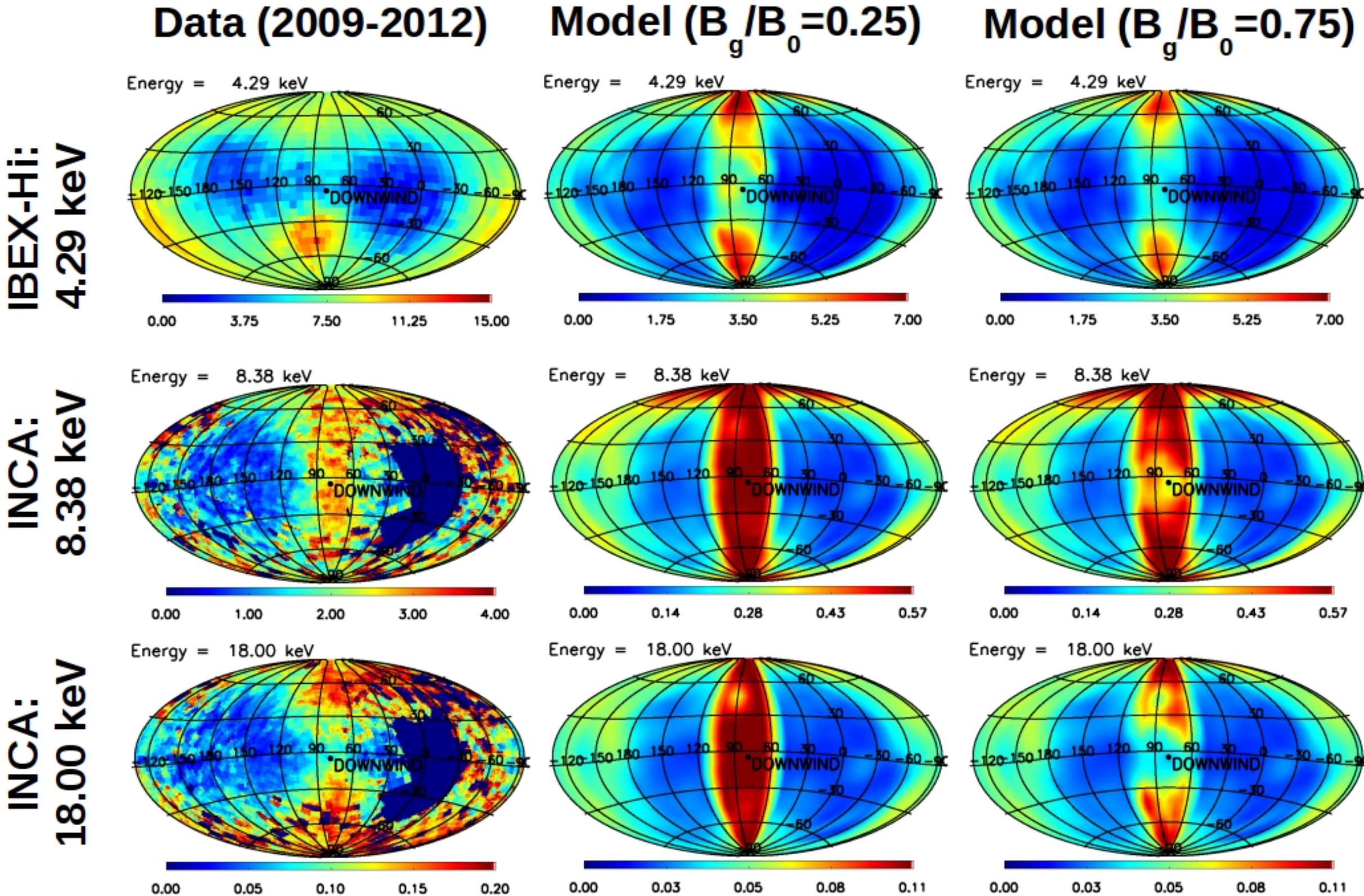

**Fig. 7:** A strong guide-field is required to replicate the energy dependence of INCA ENA observations. We present a comparison of ENA observations and model results for the 4.29, 8.38, and 18.00 keV energy bands centered on the heliotail. The 4.29 keV energy band corresponds to IBEX-Hi observations. The 8.38 and 18.00 keV energy bands correspond to INCA observations. Left column: ENA observations corresponding to an average over the years 2009-2012. Middle column: ENA model results including reconnection in the low-β region of the heliotail with a weak guide field ($B_g/B_0$=0.25). Right column: ENA model results including reconnection in the heliotail with a strong guide field ($B_g/B_0$=0.75). All maps are normalized relative to the maximum flux within each individual map to provide a qualitative comparison of the morphology. The results indicate that there is a strong guide field in the area where reconnection is occurring, as a weaker guide field cannot replicate the energy dependence of the Belt.